\providecommand{\noopsort}[1]{}
\documentclass[twocolumn]{openjournal}

\makeatletter
\expandafter\let\csname longtable*\endcsname\relax
\expandafter\let\csname endlongtable*\endcsname\relax
\makeatother

\usepackage{lipsum}
\usepackage{subcaption}

\usepackage{xcolor}
\usepackage{textgreek}
\usepackage[utf8]{inputenc}
\usepackage[english]{babel}

\usepackage{hyperref}
\hypersetup{
    unicode, 
    colorlinks=true,
    linkcolor=linkcolor,
    citecolor=linkcolor,
    filecolor=linkcolor,
    urlcolor=linkcolor,
}
\usepackage{color,colortbl}
\definecolor{linkcolor}{rgb}{0.0,0.3,0.5}
\usepackage{tensind}
\tensordelimiter{?}
\DeclareGraphicsExtensions{.bmp,.png,.jpg,.pdf}
\usepackage{verbatim}
\usepackage[normalem]{ulem}
\usepackage{orcidlink}
\usepackage{soul}

\graphicspath{ {./figs/} }

\usepackage[T1]{fontenc}

\usepackage{graphicx}	
\usepackage{amsmath, amssymb, graphicx, color}

\usepackage{xspace}

\usepackage{acro}
\DeclareAcronym{UFD}{
  short = UFD ,
  long  = ultra-faint dwarf
}
\DeclareAcronym{BHNS}{
  short = BHNS ,
  long  = black hole--neutron star
}

\DeclareAcronym{BNS}{
  short = BNS ,
  long  = binary neutron star
}

\DeclareAcronym{BBH}{
  short = BBH ,
  long  = binary black hole
}

\DeclareAcronym{NS}{
  short = NS ,
  long  = neutron star
}

\DeclareAcronym{BH}{
  short = BH ,
  long  = black hole
}

\DeclareAcronym{GW}{
  short = GW ,
  long  = gravitational wave
}

\DeclareAcronym{SN}{
  short = SN ,
  long  = supernovae
}

\DeclareAcronym{DWD}{
  short = DWD ,
  long  = double white dwarf
}

\DeclareAcronym{CE}{
  short = CE ,
  long  = common-envelope
}

\DeclareAcronym{SMT}{
  short = SMT ,
  long  = stable mass transfer
}

\DeclareAcronym{DCO}{
  short = DCO ,
  long  = double-compact object
}

\DeclareAcronym{ZAMS}{
  short = ZAMS ,
  long  = zero-age main sequence
}

\DeclareAcronym{CHE}{
  short = CHE ,
  long  = Chemically Homogeneous Evolution
}

\newcommand{\Gpcyr}{\ensuremath{\,\rm{Gpc}^{-3}\,\rm{yr}^{-1}}\xspace}

\newcommand{\UCSD}{Department of Astronomy and Astrophysics, University of California, San Diego, La Jolla, CA 92093, USA}

\begin{document}

\title{Lower Your Rates: On Claims of a Binary Black Hole Merger-Rate Crisis}

\author{Floor S. Broekgaarden\orcidlink{0000-0002-4421-4962}}
\email{fbroekgaarden@ucsd.edu}
\affiliation{\UCSD}


\begin{abstract}
Recent studies have argued that isolated binary evolution simulations generically overestimate the observed local binary black hole (BBH) merger rate, even after adopting observationally motivated variations in the metallicity-dependent cosmic star-formation history, and have interpreted this as motivation for drastic revisions to binary stellar evolution models.
We revisit these claims using a compilation of 1490 simulated BBH merger rates from 57 isolated binary-evolution studies, compared to constraints from the LIGO--Virgo--KAGRA Collaboration through GWTC-5. While $\sim$80\% of compiled submodels find rates above the GWTC-5 interval, a substantial subset reproduces or underestimates the observed rate. The literature spans several orders of magnitude, reflecting strong sensitivity to assumptions about natal kicks, common-envelope evolution, mass transfer, angular-momentum loss, remnant formation, stellar winds, initial conditions, and star-formation history.
Using 2543 pairwise BBH submodel variations constructed to isolate single physical assumptions, we identify which choices most strongly impact the simulated BBH merger rate. Low BBH merger rates are not uniquely associated with strong natal kicks or reduced low-metallicity star formation. Multiple physically motivated assumptions can independently reduce simulated rates to values consistent with observations.
We further show that simulated rates cluster into `simulation silos': frameworks producing apparent consensus within a code that does not generalize beyond it. Our results indicate that claims of a universal BBH merger-rate crisis are strongly model dependent, and underscore the importance of exploring the full parameter space across multiple population-synthesis frameworks before concluding that isolated binary evolution is in tension with gravitational-wave observations.
\end{abstract}


\maketitle

\section{Introduction}
\label{sec:intro}

The growing catalog of gravitational-wave detections from the
LIGO--Virgo--KAGRA (LVK) Collaboration has enabled increasingly precise observational constraints on the local merger-rate density of \acp{BBH} \citep{GWTC-5:catalog, GWTC-5:populations}. Recent studies have argued that isolated binary evolution simulations overestimate the observed local BBH merger rate, even after adopting observationally motivated variations in the metallicity-dependent star-formation history \citep{Sgalletta:2025, Boco:2026, Sgalletta:2026}. In particular, \citet{Boco:2026} argue that reconciling theoretical simulated rates with an observed BBH rate of $\lesssim 20\,\Gpcyr$ requires substantial revisions to binary stellar-evolution models beyond modifications to the cosmic star-formation history alone.

However, local simulated BBH merger-rates from isolated binary evolution span several orders of magnitude across the literature \citep[e.g.][]{MandelBroekgaarden:2021, Breivik:2025review, Broekgaarden:2026}. These simulated rates depend on numerous uncertain assumptions, including natal kicks, \ac{CE} evolution, mass-transfer physics, compact-object remnant formation, stellar winds, and the cosmic star formation history. It is therefore not obvious that conclusions derived from a limited set of models or parameter variations generalize across the broader landscape of isolated-binary evolution studies.

In this Letter, we therefore revisit claims of a BBH merger-rate tension by updating the compilation of isolated-binary-evolution simulated rates presented by \citet{MandelBroekgaarden:2021} and extending the analysis to include the underlying model parameter variations. We show that while many models indeed overestimate the observed local BBH merger rate, a substantial number remain consistent with current observational constraints. Moreover, we demonstrate that low BBH merger rates can arise from a much wider range of binary-evolution assumptions than  previously emphasized (i.e. only high natal BH kicks). Finally, we show that simulations can occupy distinct `simulation silos', complicating the identification of robust parameter trends within a single simulation framework. Together, these results highlight the importance of exploring the full high-dimensional parameter space before concluding that isolated binary evolution faces a BBH merger-rate crisis.

\section{Method}
\label{sec:method}

We substantially update and expand the compilation of local ($z \approx 0$) simulated merger rates from isolated binary-evolution studies presented in \citet{MandelBroekgaarden:2021} and publicly available through \citet{ZenodoReview:2021}, extending it to include studies published through 2026. The compilation includes the latest LVK constraints on the BBH, BHNS, and BNS merger rates from \citet{GWTC-5:populations, GWTC-4-pop,GWTC3,GWTC2,Abbott:2021-first-NSBH}, together with simulated rates from a broad range of isolated binary population-synthesis studies \citep{AblimitMaeda:2018,ArcaSedda:2026,Artale:2019,Baibhav:2019,Bavera:2020,Belczynski:2017,Belczynski:2020,Boco:2019,Boco:2026,Boesky:2024popsynth,Boesky:2024gw,Bouffanais:2021,Broekgaarden:2021,Broekgaarden:2022,Briel:2022,Chattaraj:2026,Chen:2025,Chruslinska:2018,Chruslinska:2019,Chu:2021,deMinkBelczynski:2015,Deng:2024,deSa:2024initial,DeSantis:2026,Dominik:2014,DorozsmaiToonen:2022,Eldridge:2019,Ghodla:2021,GiacobboMapelli:2018,GiacobboMapelli:2020,Hendriks:2023,Klencki:2018,Kruckow:2018,Lamberts:2016,Levina:2026,Li:2025,LipunovPruzhinskaya:2014,Lipunov:2016,Mapelli:2017,MapelliGiacobbo:2018,Mapelli:2020,Marinacci:2026,Mennekens:2014,Mestichelli:2025,Neijssel:2019,Olejak:2021,Olejak:2022,OShaughnessy:2009,Pellouin:2025,Rauf:2024,Riley:2020,Romagnolo:2023,Romagnolo:2025,RomanGarza:2021,Santoliquido:2020,Santoliquido:2021,Sgalletta:2025,Sgalletta:2026,Shao:2021,Smith:2026,Spera:2019,Srinivasan:2023,StevensonClarke:2022,Tang:2020,vanSon:2022,vanSon:2022-nopeaks,vanSon:2023sfrd,Xing:2024-BHNS-allZ,VignaGomez:2018,Zevin:2020,Zhu:2024}.

We include only studies that report simulated intrinsic local merger-rate  for BBHs, BHNSs, and/or BNSs and account for a metallicity-dependent cosmic star-formation history when calculating these rates. We treat each paper or distinct simulation dataset as a study and retain all reported submodels, which generally correspond to variations of one or more population-synthesis assumptions within a study. The final compilation contains 72 isolated binary-evolution studies: 57 reporting BBH rates, 39 reporting BHNS rates, and 43 reporting BNS rates.\footnote{Many studies report merger rates for more than one double compact object type, whereas others report only one.} Together, these studies provide 4194 individual (sub)model realizations, comprising 1490 BBH, 1007 BHNS, and 1697 BNS models.

For each model realization, we extract the simulated local merger-rate density 
and convert it to units of $\Gpcyr$ where necessary. We also record relevant 
information about the underlying population-synthesis calculation, including the 
code employed and the physical assumptions or parameters varied between submodels. 
This allows us to retain the full range of simulated rates reported by each study and 
to associate changes in the simulated merger rate with specific assumptions about 
binary stellar evolution. For visualization, we group submodel variations into 
broader parameter families, including, for example, {common-envelope 
efficiency}, encompassing variations in the \ac{CE} efficiency parameter $\alpha$; 
{optimistic versus pessimistic CE}, describing whether a Hertzsprung-gap 
donor initiating a CE episode is permitted to survive; {mass-transfer 
efficiency}; {compact-object remnant-mass prescriptions}; {stellar-wind 
models}; {initial conditions} (such as the initial mass function); and 
assumptions about the {cosmic star-formation and metallicity history}.

We identify pairwise `relationships' between submodels that differ in a single 
parameter or physical assumption, allowing us to quantify the corresponding change 
in merger rate. Some submodels vary more than one assumption simultaneously, so a 
single submodel may contribute to more than one parameter-family category. For 
example, consider a study that performs a $2\times2$ parameter grid varying two 
values each for $\alpha$ (CE efficiency) and $\lambda$ (envelope binding energy 
parameter). Each of the four submodels $(\alpha, \lambda) \in \{(0.5, 0.1), 
(0.5, 0.5), (1.0, 0.1), (1.0, 0.5)\}$ participates in both an $\alpha$-family 
relationship and a $\lambda$-family relationship, because it differs from at least 
one neighbour in each parameter dimension separately.

When constructing relationships within a parameter family, we apply different rules 
depending on whether the varied parameter is discrete (i.e., a labeled prescription 
or model switch) or continuous (i.e., a numerical quantity admitting a natural 
ordering). For discrete parameters:  such as varying the remnant mass prescription 
among delayed, rapid, and stochastic models,  we include all pairwise combinations: 
delayed--rapid, delayed--stochastic, and rapid--stochastic. For continuous parameters:  such as varying $\alpha \in \{0.1, 0.5, 1.0\}$,  we include only adjacent pairs 
along the ordered sequence: $\alpha = 0.1$--$0.5$ and $\alpha = 0.5$--$1.0$, but not 
the non-adjacent pair $\alpha = 0.1$--$1.0$. This reflects the principle that for 
ordered numerical parameters, intermediate values already capture the transition, 
whereas for discrete prescription switches there is no inherent ordering and all 
pairings are equally informative. As a consequence of these rules, a study with a 
large discrete parameter grid will often contribute more relationships than submodels, which  explains why the total relationship counts ($2543$ BH--BH, $3535$ NS--NS, and $1863$ NS--BH) exceed the number of submodels in each category. See our supplementary material for full details.

Our goal is not to perform a homogeneous reanalysis of the underlying simulations, but to characterize the diversity of simulated merger-rate across the isolated-binary literature and investigate which assumptions (most strongly) affect the simulated  local BBH merger rate. Throughout this work, we consider only isolated binary-evolution models; simulated rates from other formation channels are discussed in \citet{MandelBroekgaarden:2021} and are not included here.

Full details of the literature compilation, merger-rate conversions, recorded information,  submodel parameter assumptions, and parameter classifications are provided in the \href{https://floorbroekgaarden.github.io/lower-your-rates/}{\textbf{Supplementary Material}}. To facilitate reproducibility, all underlying data, supplementary tables, and code used to generate the figures are publicly available through Zenodo \citep{BroekgaardenZenodo:2026-Lower-Your-Rates}. The compilation can also be explored through an interactive online repository at \href{https://floorbroekgaarden.github.io/lower-your-rates/}{https://floorbroekgaarden.github.io/lower-your-rates/}.

\begin{figure*}
\centering
\includegraphics[width=1\linewidth]{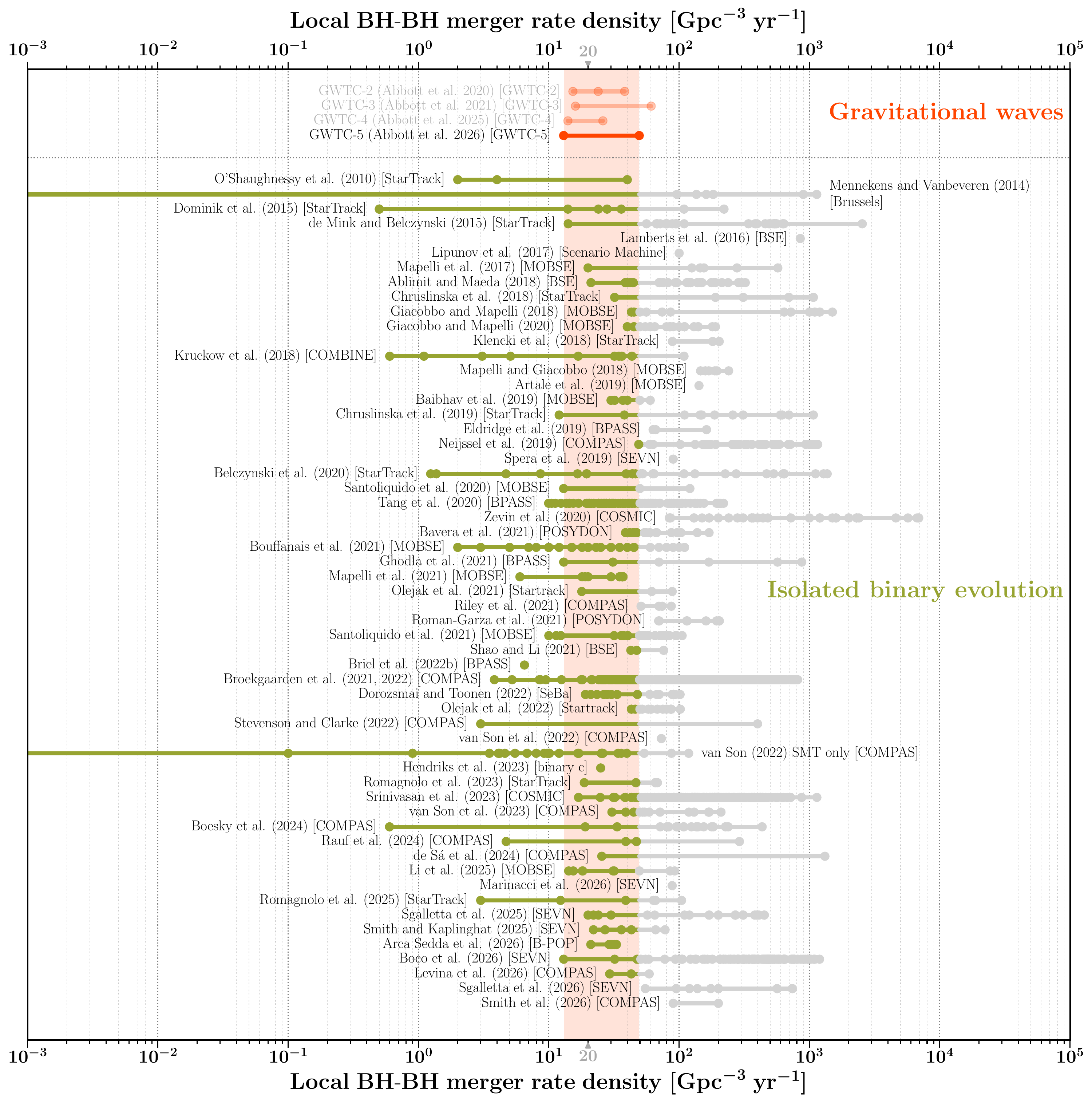}
\caption{
Compilation of simulated local BBH merger-rate from isolated binary evolution studies compared to the observational constraints from GWTC-5 (and earlier LVK constraints).
While many isolated-binary models find merger rates above the observed LVK range (gray points), a substantial number of simulations reproduce or underestimate the observed local BBH merger rate (green points).
The simulated rates span several orders of magnitude across the literature, highlighting the strong dependence of simulated merger rates on assumptions regarding massive star binary evolution  and the cosmic star formation history modeling.
These results demonstrate that claims of a universal tension between isolated binary evolution and the observed BBH merger rate are presently highly model dependent.
Each horizontal line represents the BBH range of one study with (where present) scatter points indicating individual submodel rates from that simulation. The author of the study and code are annotated in text next to the BBH rate range.
We annotate the rate of $20\Gpcyr$ which is used as the observed rate in \citet{Boco:2026}.
For the GWTC-5 BBH range we use the interval of $27.5$--$49.4$\Gpcyr at $z\approx0.2$ \citep[joint rates,][Table~2]{GWTC-5:populations}, extending down to $\approx13$~\Gpcyr at $z\approx0$ (Figure~5 therein).
}
\label{fig:BBH-isolated-review-rates}
\end{figure*}

\section{Results}
\label{sec:results}

\subsection{The literature does not support a universal BBH merger-rate crisis}
\label{sec:results-literature}

Figure~\ref{fig:BBH-isolated-review-rates} compares published simulated local BBH merger rates from isolated binary evolution with successive LVK constraints. The BHNS and BNS rates are in Appendix~\ref{ap:ap-BHNS-and-BNS-parameter-variations}.

The simulated local BBH merger rates span several orders of magnitude across the isolated binary evolution literature. Although many models find BBH merger rates substantially above the observed LVK range, Figure~\ref{fig:BBH-isolated-review-rates} demonstrates that these models do not represent the full range of isolated-binary evolution model simulated rates. The literature also contains a substantial number of simulations that reproduce or underestimate the observed local BBH merger rate.

Nevertheless, the literature is strongly weighted toward high merger rates: approximately ($\sim80\%$) of the 1490 compiled BBH submodels find rates above the upper boundary of the current GWTC-5 interval. This is consistent with recent studies that have identified substantial BBH overproduction within particular model frameworks \citep{Sgalletta:2025, Boco:2026}. For example, most BBH submodels presented by \citet{Sgalletta:2025} lie above the GWTC-5 upper boundary (and the few BBH rates below $20\Gpcyr$)  correspond to models in which BHs receive large natal kicks drawn from the distribution of \citet{Hobbs:2005}.) A small number of studies report only BBH merger rates above the observational interval across all of their submodels \citep[e.g.,][]{Eldridge:2019,Zevin:2020,RomanGarza:2020,Marinacci:2026,Sgalletta:2026,Smith:2026}.

However, these high-rate studies do not represent the full range of simulated isolated-binary rates. Numerous studies contain at least one submodel within the GWTC-5 interval, and many also contain many submodels below it. The relevant distinction is therefore between a literature distribution that is \emph{skewed toward} high rates and a theoretical framework that \emph{inevitably} finds high simulation rates. Figure~\ref{fig:BBH-isolated-review-rates} supports the former statement but not the latter. 

We further caution that the individual simulated merger rates in this compilation should not be interpreted as independent draws from the space of plausible isolated-binary models. Submodels within a given study typically share the same population-synthesis code and many of the same physical assumptions, while some model frameworks contribute substantially more submodels than others. Consequently, the literature distribution is weighted toward the behavior of the most extensively sampled frameworks, several of which systematically find relatively high BBH merger rates. We examine this clustering into distinct `simulation silos' in Section~~\ref{sec:results-silos}). 

We also note that the apparent severity of the discrepancy also depends on the observational benchmark. \citet{Boco:2026} quantify overproduction relative to a GWTC-4 rate of approximately $20\Gpcyr$, which comes from the central value of the GWTC-4 inferred BBH rate \citet{GWTC-4-pop}. 
The latest GWTC-5 analysis instead gives an interval of approximately $27.5$--$49.4$~\Gpcyr at $z\approx0.2$ \citep[Table~2, joint rates][]{GWTC-5:populations}, extending down to $\approx13$~\Gpcyr\ at $z\approx0$ (Figure~5 \citealt[][]{GWTC-5:populations}); the higher rate range of $49.4$\Gpcyr relaxes the tension.
For a fixed simulated rate, using the upper boundary of the GWTC-5 interval rather than 20\Gpcyr reduces the inferred overproduction factor already by approximately a factor up to  $2.5$. Thus, although many published models remain above current observational constraints, the quantitative level of tension has decreased as the observationally inferred rate has evolved.

Taken together, these results do not support a universal conclusion that isolated binary evolution generically overestimates the local BBH merger rate. They instead show that the existence and magnitude of the tension depend strongly on the adopted stellar- and binary-evolution assumptions, star formation history assumptions, the population-synthesis framework, and the observational reference value.

\begin{figure}
    \centering
    \includegraphics[width=1\columnwidth]{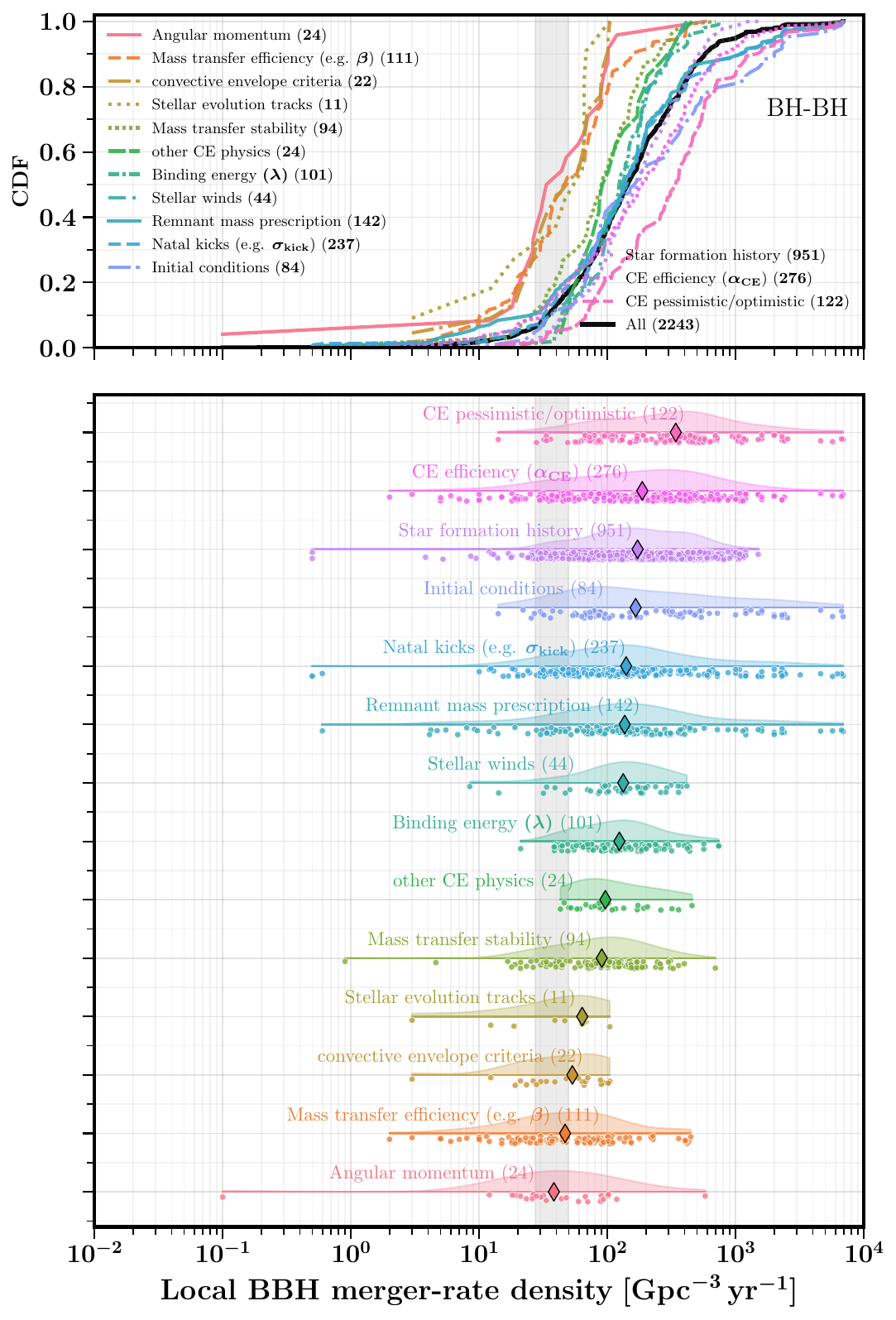}
    \caption{Distribution of local simulated BBH merger-rates grouped by the uncertain parameter family varied in each study. The upper panel shows the cumulative distribution function (CDF) of published simulated merger rates for each parameter family, while the lower panel shows the corresponding distributions and individual model realizations (scatter points that are given a small y offset for visualization purposes). Shaded regions indicate the current GWTC-5 merger-rate interval. We only show parameter families with at least 10 datapoints. Nearly all parameter families contain models both consistent and inconsistent with current observational constraints. The substantial overlap between distributions demonstrates that low BBH merger rates are not uniquely associated with natal kicks or metallicity evolution (star formation history model), but can arise from a broad range of assumptions regarding binary stellar evolution.}
    \label{fig:parameter-variation-violin}
\end{figure}

\begin{figure}
    \centering
    \includegraphics[width=1\columnwidth]{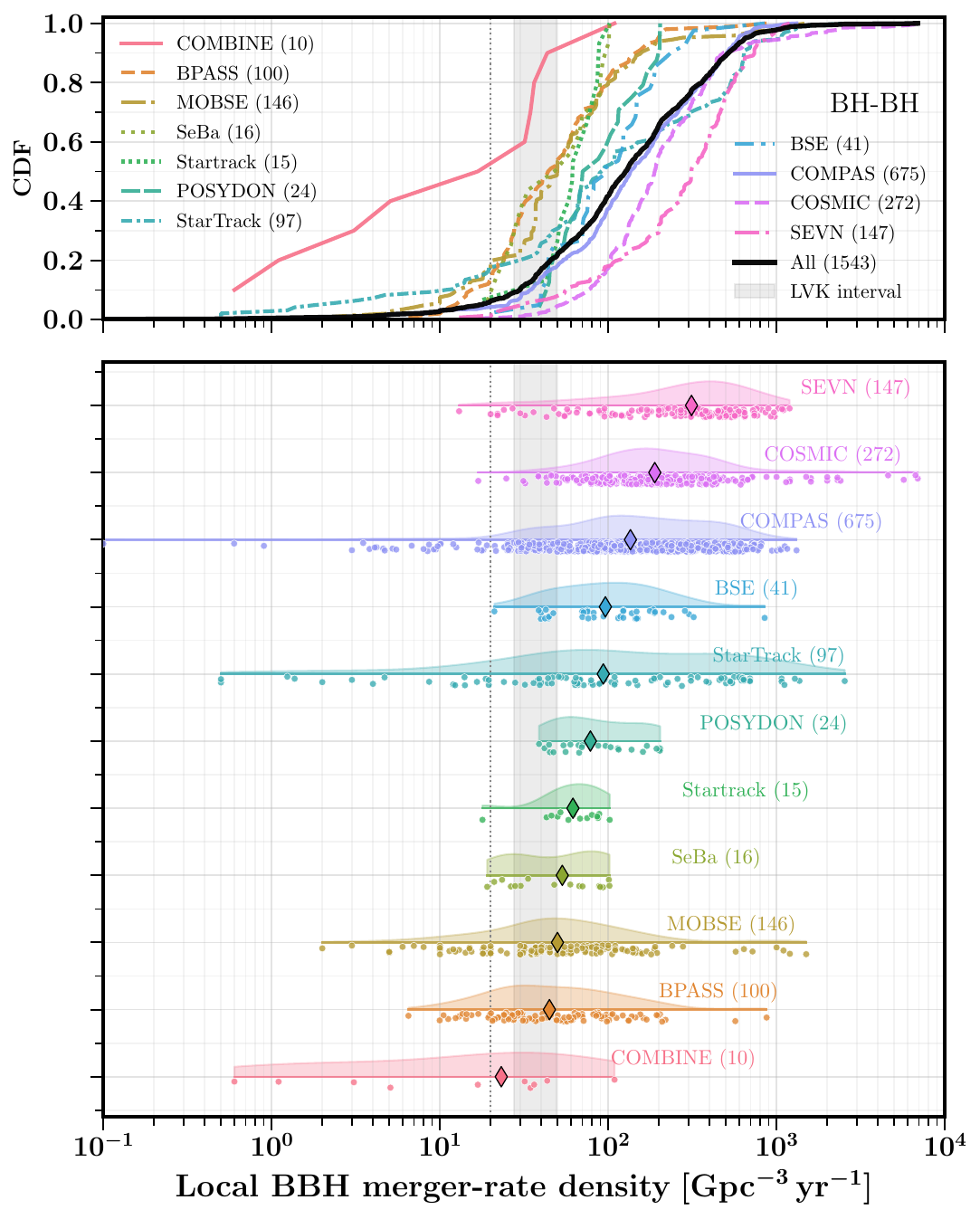}
    \caption{Same as Figure~\ref{fig:parameter-variation-violin}, but grouping simulated merger rates by population-synthesis framework rather than parameter family. Different codes populate somewhat different regions of merger-rate space, reflecting differences in their underlying implementations of stellar and binary evolution. At the same time, substantial overlap exists among the simulated merger rate distributions. This clustering by framework illustrates the importance of simulation silos: conclusions drawn from a single population-synthesis code may not generalize across the broader landscape of isolated-binary evolution models.}
    \label{fig:violin-population-synthesis-code}
\end{figure}

\subsection{Multiple physical assumptions can reconcile BBH rates with observations}
\label{sec:results-parameter-impact-COC-rates}

Recent work has emphasized two mechanisms for reducing simulated BBH merger rates: large BH natal kicks and cosmic star-formation histories containing less low-metallicity star formation \citep{Sgalletta:2025,Boco:2026}. We therefore use the broader literature compilation to examine whether low BBH rates are confined to these assumptions.

Figure~\ref{fig:parameter-variation-violin} groups the published  simulated BBH merger rates from the literature according to the physical parameter family varied within each study. The figure summarizes the distribution of the compiled published realizations (i.e. from Figure~\ref{fig:BBH-isolated-review-rates}); it should not be interpreted as a prior-weighted probability distribution over physically plausible models. Moreover, because some submodels vary more than one assumption, a single submodel may contribute to more than one parameter-family category (see Section~\ref{sec:method} for details). 

Several parameter families span more than two orders of magnitude in local BBH merger rate. Particularly broad ranges occur among studies varying common-envelope evolution physics, metallicity-dependent cosmic star-formation history, initial conditions, natal kicks, remnant-mass prescription, and angular-momentum loss. Within these families, published simulation BBH merger rates extend from below 20\Gpcyr to values exceeding the GWTC-5 upper limit by more than an order of magnitude.

The most important result is that low BBH rates are not confined to natal-kick or star formation history variations. Models within or below the GWTC-5 interval also occur when varying other parameters including common-envelope efficiency and survival, envelope binding energies, mass-transfer stability, mass-transfer efficiency, angular-momentum loss, remnant formation, stellar-evolution tracks, and stellar winds. The literature therefore contains multiple physically distinct routes by which the simulated BBH merger rate can be reduced.

The table shown in Figure~\ref{tab:examples-rate-reductions} illustrates this result by zooming in on some of the BBH relationships that show most change in the BBH rate. In each example, changing a particular assumption moves the simulated BBH rate significantly, often from above the GWTC-5 interval to within or below it while retaining the remainder of the study's modeling framework as closely as possible. These examples provide a more controlled complement to the pooled distributions in Figure~\ref{fig:parameter-variation-violin}. They demonstrate that substantial reductions can result indeed from natal kicks or star formation history assumption variations, but also from changes in remnant mass prescription, mass transfer stability, initial conditions, and CE assumptions (among other variations).

These results do not imply that all low-rate models are equally plausible. Some may conflict with BBH mass, spin, or redshift distributions, electromagnetic observations, or independent constraints on stellar and binary evolution. They do show, however, that the local BBH merger rate alone does not uniquely select strong natal kicks, suppressed low-metallicity star formation, or any other single physical prescription. Agreement with the observed rate can arise through several distinct and potentially compensating combinations of assumptions.

\subsection{Simulation silos can make model-dependent results appear universal}
\label{sec:results-silos}

Figure~\ref{fig:violin-population-synthesis-code} groups the compiled BBH simulation rates by population-synthesis framework. A striking feature is that we find that the resulting distributions cluster by code: some frameworks predominantly populate high-rate regions, whereas others include a larger proportion of models within or below the GWTC-5 interval. In particular, SEVN, COSMIC, and COMPAS contribute many of the highest simulated BBH rates, while also accounting for the largest numbers of submodels in the compilation. The high-rate models discussed by \citet{Boco:2026} and \citet{Sgalletta:2025} are based on SEVN simulations, which occupy the upper end of the distributions shown here. Thus, conclusions drawn from this subset of models may partly reflect the characteristic rate distribution of the underlying population-synthesis framework rather than a generic feature of isolated binary evolution.

The code-level differences in Figure~\ref{fig:violin-population-synthesis-code} should not be interpreted as the isolated effect of numerical implementation. A population-synthesis framework comprises a collection of choices concerning stellar-evolution tracks, mass-transfer stability, common-envelope evolution, stellar winds, remnant formation, natal kicks, initial (zero-age-main-sequence) properties, and cosmic star-formation history. Furthermore, individual studies generally vary only a subset of these assumptions while holding the remainder fixed. The distributions in Figure~\ref{fig:violin-population-synthesis-code} therefore reflect both differences among codes and differences in the regions of parameter space explored with each code.

We refer to this concentration of studies within restricted combinations of code and assumptions as \emph{simulation silos}. A set of studies using closely related prescriptions may repeatedly obtain high merger rates, producing an apparent consensus within that region of model space. Such agreement demonstrates robustness conditional on the shared assumptions, but does not by itself establish that the conclusion generalizes to isolated binary evolution as a whole.

The substantial overlap among the code distributions reinforces this point. Similar merger rates can be produced by different frameworks, while a single framework can produce widely different rates when its assumptions are varied. The local BBH merger rate is therefore a highly degenerate observable: it constrains combinations of assumptions rather than uniquely identifying the physical prescription responsible for agreement or disagreement with observations.

Figures~\ref{fig:parameter-variation-violin} and \ref{fig:violin-population-synthesis-code} together show why conclusions drawn from one code or a narrow set of parameter variations can appear more general than they are. The current BBH merger-rate tension is best understood not as a universal failure of isolated binary evolution, but as a tension affecting particular regions of a large and incompletely explored model space, motivating future studies to explore model outcomes in comparison with other observables such as BBH mass distributions, spins, and redshift evolution.

\subsection{Extension to BHNS and BNS merger rates}
\label{sec:results-other-dcos}

Appendix~\ref{ap:ap-BHNS-and-BNS-parameter-variations} presents the corresponding parameter-family and population-synthesis-code distributions for BHNS and BNS merger rates. Similar to BBHs, the simulated merger rates span broad ranges and exhibit substantial clustering by simulation framework. Many parameter families produce variations of several orders of magnitude, cautioning against attributing agreement or disagreement with observational constraints to any single physical assumption based on results from only one modeling framework.

The parameter families and codes associated with the broadest or highest rate distributions also differ among compact-object populations. For example, the highest BHNS merger rates are predominantly contributed by BPASS, SEVN, and COMPAS models, whereas the highest BNS rates are primarily associated with COSMIC and BPASS. This population-dependent behavior further illustrates that trends identified for BBHs do not necessarily generalize to BHNSs or BNSs. Nevertheless, the majority of the compiled BHNS and BNS simulated merger rates lie within the corresponding merger-rate intervals inferred from GWTC-5 \citep{GWTC-5:populations}.

\section{Conclusions}
\label{sec:conclusions}

We compiled and updated local merger-rate estimates from isolated binary-evolution studies and compared them with the latest observational constraints from the LVK Collaboration. By investigating the individual submodels reported within each study, we characterize both the overall diversity of simulated rates and the assumptions responsible for large variations.

Our main conclusions are:

\begin{itemize}

\item The literature distribution of simulated BBH merger rates is strongly skewed toward high values: many isolated-binary models yield local rates above the current GWTC-5 interval. However, the literature also contains numerous submodels that reproduce or underestimate the observed rate. The available simulations therefore do not support the stronger claim that isolated binary evolution inevitably overproduces BBH mergers.  

\item Simulated BBH merger rates span several orders of magnitude across the literature. Large variations arise not only from natal kicks and the metallicity-dependent cosmic star-formation history, but also from assumptions concerning common-envelope evolution, mass-transfer stability and efficiency, angular-momentum loss, remnant formation, stellar winds, stellar-evolution tracks, and assumed initial (zero-age-main-sequence) properties.

\item Low BBH merger rates are therefore not uniquely associated with any single physical prescription. Multiple, physically distinct changes to the isolated binary evolution population synthesis model can shift simulated rates from above the GWTC-5 interval to within or below it. The local merger rate alone consequently constrains combinations of assumptions rather than uniquely identifying the physics responsible for agreement with observations.\\

\item Simulated merger rates cluster by population-synthesis framework. Because studies using the same code generally share many assumptions and explore only restricted regions of parameter space, repeated agreement within one framework can create an apparent consensus that does not generalize across isolated binary evolution as a whole. We refer to these restricted regions of model space as ``simulation silos.''

\item The individual submodels in the literature should therefore not be interpreted as independent or prior-weighted samples of the full space of plausible isolated binary-evolution models. Frameworks that contribute many submodels can disproportionately shape the apparent literature distribution, particularly when those frameworks systematically favor high merger rates.

\item Robustly assessing a possible tension between isolated binary evolution and gravitational-wave observations will require broader and more coordinated exploration of the correlated, high-dimensional model space across multiple population-synthesis frameworks. Merger rates should also be considered jointly with additional observables, including the mass, spin, and redshift distributions of merging binaries and independent electromagnetic constraints on massive-star and binary evolution.

\end{itemize}

Taken together, our results do not support a universal BBH merger-rate crisis for isolated binary evolution. Instead, the existence and magnitude of the tension depend strongly on the adopted population-synthesis framework, the region of parameter space explored, the cosmic star-formation and metallicity history, and the observational benchmark used for comparison. Claims of a generic failure of isolated binary evolution are therefore premature without a substantially broader exploration of the underlying model space.

\section*{Acknowledgments}
FSB acknowledges support from NASA HPOSS grant 80NSSC25K7555 under award number 316592-0000 as well as support from an NSF 22-624 Astronomy and Astrophysics Research Grants grant under award number 2606407. This work benefited from many of the discussions FSB had with members from the UCSD Gravitational Wave Paleontology Lab. in particular: Sasha Levina, Steffani Grondin, Melanie Santiago, Julia Haynes, Kyle Rocha, Tyler Smith, Rhea Kumar, Kera Yu and Joseph Rodriguez. 

\section*{Software Acknowledgement}

This work made use of the following software packages: \textsc{astropy} \citep{astropy:2013,astropy:2018,astropy:2022,astropy_17756022}, \textsc{Jupyter} \citep{2007CSE.....9c..21P,kluyver2016jupyter}, \textsc{matplotlib} \citep{Hunter:2007}, \textsc{numpy} \citep{numpy}, \textsc{python} \citep{python}, \textsc{scipy} \citep{2020SciPy-NMeth, scipy_14880408},  \textsc{PyTables} \citep{pytables}, and WebPlotDigitizer \citep{rohatgi2020webplotdigitizer}.  Claude code and ChatGPT were used to help with converting our static figures using \texttt{plotly} \citep{plotly, plotly_14366349} into interactive figures to help with data visualization and for cleaning up and documenting code.  
This research has made use of NASA's Astrophysics Data System.
Software citation information aggregated using \texttt{\href{https://www.tomwagg.com/software-citation-station/}{The Software Citation Station}} \citep{software-citation-station-paper,software-citation-station-zenodo}.

\section*{Data Availability}
All underlying data, supplementary tables, and code used to generate the figures are publicly available through Zenodo \citep{BroekgaardenZenodo:2026-Lower-Your-Rates}. The data and figures can also be explored through an interactive online repository at \href{https://floorbroekgaarden.github.io/lower-your-rates/}{https://floorbroekgaarden.github.io/lower-your-rates/}.

This work made use of publicly available data and code from \citep{MandelBroekgaarden:2021, ZenodoReview:2021} for creating the BBH, BHNS, and BNS rates compilation. 

\section*{Author contributions}
FSB led the research and writing of the manuscript work.

\bibliographystyle{mnras}
\bibliography{lower-your-rates}

@ARTICLE{2007CSE.....9c..21P,
       author = {{Perez}, Fernando and {Granger}, Brian E.},
        title = "{IPython: A System for Interactive Scientific Computing}",
      journal = {Computing in Science and Engineering},
         year = "2007",
        month = "Jan",
       volume = {9},
       number = {3},
        pages = {21-29},
          doi = {10.1109/MCSE.2007.53},
       adsurl = {https://ui.adsabs.harvard.edu/abs/2007CSE.....9c..21P}
}

@ARTICLE{Broekgaarden:2026,
       author = {{Broekgaarden}, Floor S. and {Lam}, Ana and {Levina}, Sasha and {Klencki}, Jakub and {Rocha}, Kyle A. and {van Son}, Lieke and {Grondin}, Steffani M. and {Gallegos-Garcia}, Monica and {Metzger}, Brian D. and {Ramirez-Ruiz}, Enrico and {Twum}, Angela and {Santiago}, Melanie and {Haynes}, Julia and {Smith}, Tyler B. and {Romagnolo}, Amedeo and {Berger}, Edo and {de S{\'a}}, Lucas M.},
        title = "{How Common Are Common Envelopes? Quantifying Their Role in Forming Gravitational-Wave Sources}",
      journal = {arXiv e-prints},
         year = 2026,
        month = jun,
          eid = {arXiv:2606.05322},
        pages = {arXiv:2606.05322},
          doi = {10.48550/arXiv.2606.05322},
archivePrefix = {arXiv},
       eprint = {2606.05322},
 primaryClass = {astro-ph.HE},
       adsurl = {https://ui.adsabs.harvard.edu/abs/2026arXiv260605322B}
}

@article{2020SciPy-NMeth,
  author        = {Virtanen, Pauli and Gommers, Ralf and Oliphant, Travis E. and Haberland, Matt and Reddy, Tyler and Cournapeau, David and Burovski, Evgeni and Peterson, Pearu and Weckesser, Warren and Bright, Jonathan and {van der Walt}, St{\'e}fan J. and Brett, Matthew and Wilson, Joshua and Millman, K. Jarrod and Mayorov, Nikolay and Nelson, Andrew R. J. and Jones, Eric and Kern, Robert and Larson, Eric and Carey, C J and Polat, {\.I}lhan and Feng, Yu and Moore, Eric W. and {VanderPlas}, Jake and Laxalde, Denis and Perktold, Josef and Cimrman, Robert and Henriksen, Ian and Quintero, E. A. and Harris, Charles R. and Archibald, Anne M. and Ribeiro, Ant{\^o}nio H. and Pedregosa, Fabian and {van Mulbregt}, Paul and {SciPy 1.0 Contributors}},
  title         = {{{SciPy} 1.0: Fundamental Algorithms for Scientific Computing in Python}},
  journal       = {Nature Methods},
  year          = {2020},
  volume        = {17},
  pages         = {261--272},
  adsurl        = {https://rdcu.be/b08Wh},
  doi           = {10.1038/s41592-019-0686-2}
}

@ARTICLE{Abbott:2021-first-NSBH,
       author = {{Abbott}, R. and {Abbott}, T.~D. and {Abraham}, S. and {Acernese}, F. and {Ackley}, K. and {Adams}, A. and {Adams}, C. and {Adhikari}, R.~X. and {Adya}, V.~B. and {Affeldt}, C. and {Agarwal}, D. and {Agathos}, M. and {Agatsuma}, K. and {Aggarwal}, N. and {Aguiar}, O.~D. and {Aiello}, L. and {Ain}, A. and {Ajith}, P. and {Akutsu}, T. and {Aleman}, K.~M. and {Allen}, G. and {Allocca}, A. and {Altin}, P.~A. and {Amato}, A. and {Anand}, S. and {Ananyeva}, A. and {Zlochower}, Y. and {Zucker}, M.~E. and {Zweizig}, J. and {Ligo Scientific Collaboration} and {VIRGO Collaboration} and {KAGRA Collaboration}},
        title = "{Observation of Gravitational Waves from Two Neutron Star-Black Hole Coalescences}",
      journal = {\apjl},
         year = 2021,
        month = jul,
       volume = {915},
       number = {1},
          eid = {L5},
        pages = {L5},
          doi = {10.3847/2041-8213/ac082e},
archivePrefix = {arXiv},
       eprint = {2106.15163},
 primaryClass = {astro-ph.HE},
       adsurl = {https://ui.adsabs.harvard.edu/abs/2021ApJ...915L...5A}
}

@ARTICLE{AblimitMaeda:2018,
       author = {{Ablimit}, Iminhaji and {Maeda}, Keiichi},
        title = "{Monte Carlo Population Synthesis on Massive Star Binaries: Astrophysical Implications for Gravitational-wave Sources}",
      journal = {\apj},
         year = 2018,
        month = oct,
       volume = {866},
       number = {2},
          eid = {151},
        pages = {151},
          doi = {10.3847/1538-4357/aae378},
archivePrefix = {arXiv},
       eprint = {1710.05504},
 primaryClass = {astro-ph.HE},
       adsurl = {https://ui.adsabs.harvard.edu/abs/2018ApJ...866..151A}
}

@ARTICLE{ArcaSedda:2026,
       author = {{Arca Sedda}, Manuel and {Paiella}, Lavinia and {Ugolini}, Cristiano and {Santoliquido}, Filippo and {Mestichelli}, Benedetta and {Usai}, Ilaria and {Simonato}, Filippo and {Branchesi}, Marica},
        title = "{Isolated or Dynamical? Tracing Black Hole Binary Formation through the Population of Gravitational-Wave Sources}",
      journal = {arXiv e-prints},
         year = 2026,
        month = mar,
          eid = {arXiv:2603.20430},
        pages = {arXiv:2603.20430},
          doi = {10.48550/arXiv.2603.20430},
archivePrefix = {arXiv},
       eprint = {2603.20430},
 primaryClass = {astro-ph.GA},
       adsurl = {https://ui.adsabs.harvard.edu/abs/2026arXiv260320430A}
}

@ARTICLE{Artale:2019,
       author = {{Artale}, M. Celeste and {Mapelli}, Michela and {Giacobbo}, Nicola and
         {Sabha}, Nadeen B. and {Spera}, Mario and {Santoliquido}, Filippo and
         {Bressan}, Alessandro},
        title = "{Host galaxies of merging compact objects: mass, star formation rate, metallicity, and colours}",
      journal = {\mnras},
         year = 2019,
        month = aug,
       volume = {487},
       number = {2},
        pages = {1675-1688},
          doi = {10.1093/mnras/stz1382},
archivePrefix = {arXiv},
       eprint = {1903.00083},
 primaryClass = {astro-ph.GA},
       adsurl = {https://ui.adsabs.harvard.edu/abs/2019MNRAS.487.1675A}
}

@article{astropy:2013,
Adsurl = {http://adsabs.harvard.edu/abs/2013A%26A...558A..33A},
Archiveprefix = {arXiv},
Author = {{Astropy Collaboration} and {Robitaille}, T.~P. and {Tollerud}, E.~J. and {Greenfield}, P. and {Droettboom}, M. and {Bray}, E. and {Aldcroft}, T. and {Davis}, M. and {Ginsburg}, A. and {Price-Whelan}, A.~M. and {Kerzendorf}, W.~E. and {Conley}, A. and {Crighton}, N. and {Barbary}, K. and {Muna}, D. and {Ferguson}, H. and {Grollier}, F. and {Parikh}, M.~M. and {Nair}, P.~H. and {Unther}, H.~M. and {Deil}, C. and {Woillez}, J. and {Conseil}, S. and {Kramer}, R. and {Turner}, J.~E.~H. and {Singer}, L. and {Fox}, R. and {Weaver}, B.~A. and {Zabalza}, V. and {Edwards}, Z.~I. and {Azalee Bostroem}, K. and {Burke}, D.~J. and {Casey}, A.~R. and {Crawford}, S.~M. and {Dencheva}, N. and {Ely}, J. and {Jenness}, T. and {Labrie}, K. and {Lim}, P.~L. and {Pierfederici}, F. and {Pontzen}, A. and {Ptak}, A. and {Refsdal}, B. and {Servillat}, M. and {Streicher}, O.},
Doi = {10.1051/0004-6361/201322068},
Eid = {A33},
Eprint = {1307.6212},
Journal = {\aap},
Month = oct,
Pages = {A33},
Primaryclass = {astro-ph.IM},
Title = {{Astropy: A community Python package for astronomy}},
Volume = 558,
Year = 2013}

@ARTICLE{astropy:2018,
       author = {{Astropy Collaboration} and {Price-Whelan}, A.~M. and
         {Sip{\H{o}}cz}, B.~M. and {G{\"u}nther}, H.~M. and {Lim}, P.~L. and
         {Crawford}, S.~M. and {Conseil}, S. and {Shupe}, D.~L. and
         {Craig}, M.~W. and {Dencheva}, N. and {Ginsburg}, A. and {Vand
        erPlas}, J.~T. and {Bradley}, L.~D. and {P{\'e}rez-Su{\'a}rez}, D. and
         {de Val-Borro}, M. and {Aldcroft}, T.~L. and {Cruz}, K.~L. and
         {Robitaille}, T.~P. and {Tollerud}, E.~J. and {Ardelean}, C. and
         {Babej}, T. and {Bach}, Y.~P. and {Bachetti}, M. and {Bakanov}, A.~V. and
         {Bamford}, S.~P. and {Barentsen}, G. and {Barmby}, P. and
         {Baumbach}, A. and {Berry}, K.~L. and {Biscani}, F. and {Boquien}, M. and
         {Bostroem}, K.~A. and {Bouma}, L.~G. and {Brammer}, G.~B. and
         {Bray}, E.~M. and {Breytenbach}, H. and {Buddelmeijer}, H. and
         {Burke}, D.~J. and {Calderone}, G. and {Cano Rodr{\'\i}guez}, J.~L. and
         {Cara}, M. and {Cardoso}, J.~V.~M. and {Cheedella}, S. and {Copin}, Y. and
         {Corrales}, L. and {Crichton}, D. and {D'Avella}, D. and {Deil}, C. and
         {Depagne}, {\'E}. and {Dietrich}, J.~P. and {Donath}, A. and
         {Droettboom}, M. and {Earl}, N. and {Erben}, T. and {Fabbro}, S. and
         {Ferreira}, L.~A. and {Finethy}, T. and {Fox}, R.~T. and
         {Garrison}, L.~H. and {Gibbons}, S.~L.~J. and {Goldstein}, D.~A. and
         {Gommers}, R. and {Greco}, J.~P. and {Greenfield}, P. and
         {Groener}, A.~M. and {Grollier}, F. and {Hagen}, A. and {Hirst}, P. and
         {Homeier}, D. and {Horton}, A.~J. and {Hosseinzadeh}, G. and {Hu}, L. and
         {Hunkeler}, J.~S. and {Ivezi{\'c}}, {\v{Z}}. and {Jain}, A. and
         {Jenness}, T. and {Kanarek}, G. and {Kendrew}, S. and {Kern}, N.~S. and
         {Kerzendorf}, W.~E. and {Khvalko}, A. and {King}, J. and {Kirkby}, D. and
         {Kulkarni}, A.~M. and {Kumar}, A. and {Lee}, A. and {Lenz}, D. and
         {Littlefair}, S.~P. and {Ma}, Z. and {Macleod}, D.~M. and
         {Mastropietro}, M. and {McCully}, C. and {Montagnac}, S. and
         {Morris}, B.~M. and {Mueller}, M. and {Mumford}, S.~J. and {Muna}, D. and
         {Murphy}, N.~A. and {Nelson}, S. and {Nguyen}, G.~H. and
         {Ninan}, J.~P. and {N{\"o}the}, M. and {Ogaz}, S. and {Oh}, S. and
         {Parejko}, J.~K. and {Parley}, N. and {Pascual}, S. and {Patil}, R. and
         {Patil}, A.~A. and {Plunkett}, A.~L. and {Prochaska}, J.~X. and
         {Rastogi}, T. and {Reddy Janga}, V. and {Sabater}, J. and
         {Sakurikar}, P. and {Seifert}, M. and {Sherbert}, L.~E. and
         {Sherwood-Taylor}, H. and {Shih}, A.~Y. and {Sick}, J. and
         {Silbiger}, M.~T. and {Singanamalla}, S. and {Singer}, L.~P. and
         {Sladen}, P.~H. and {Sooley}, K.~A. and {Sornarajah}, S. and
         {Streicher}, O. and {Teuben}, P. and {Thomas}, S.~W. and
         {Tremblay}, G.~R. and {Turner}, J.~E.~H. and {Terr{\'o}n}, V. and
         {van Kerkwijk}, M.~H. and {de la Vega}, A. and {Watkins}, L.~L. and
         {Weaver}, B.~A. and {Whitmore}, J.~B. and {Woillez}, J. and
         {Zabalza}, V. and {Astropy Contributors}},
        title = "{The Astropy Project: Building an Open-science Project and Status of the v2.0 Core Package}",
      journal = {\aj},
         year = 2018,
        month = sep,
       volume = {156},
       number = {3},
          eid = {123},
        pages = {123},
          doi = {10.3847/1538-3881/aabc4f},
archivePrefix = {arXiv},
       eprint = {1801.02634},
 primaryClass = {astro-ph.IM},
       adsurl = {https://ui.adsabs.harvard.edu/abs/2018AJ....156..123A}
}

@ARTICLE{astropy:2022,
       author = {{Astropy Collaboration} and {Price-Whelan}, Adrian M. and {Lim}, Pey Lian and {Earl}, Nicholas and {Starkman}, Nathaniel and {Bradley}, Larry and {Shupe}, David L. and {Patil}, Aarya A. and {Corrales}, Lia and {Brasseur}, C.~E. and {N{"o}the}, Maximilian and {Donath}, Axel and {Tollerud}, Erik and {Morris}, Brett M. and {Ginsburg}, Adam and {Vaher}, Eero and {Weaver}, Benjamin A. and {Tocknell}, James and {Jamieson}, William and {van Kerkwijk}, Marten H. and {Robitaille}, Thomas P. and {Merry}, Bruce and {Bachetti}, Matteo and {G{"u}nther}, H. Moritz and {Aldcroft}, Thomas L. and {Alvarado-Montes}, Jaime A. and {Archibald}, Anne M. and {B{'o}di}, Attila and {Bapat}, Shreyas and {Barentsen}, Geert and {Baz{'a}n}, Juanjo and {Biswas}, Manish and {Boquien}, M{'e}d{'e}ric and {Burke}, D.~J. and {Cara}, Daria and {Cara}, Mihai and {Conroy}, Kyle E. and {Conseil}, Simon and {Craig}, Matthew W. and {Cross}, Robert M. and {Cruz}, Kelle L. and {D'Eugenio}, Francesco and {Dencheva}, Nadia and {Devillepoix}, Hadrien A.~R. and {Dietrich}, J{"o}rg P. and {Eigenbrot}, Arthur Davis and {Erben}, Thomas and {Ferreira}, Leonardo and {Foreman-Mackey}, Daniel and {Fox}, Ryan and {Freij}, Nabil and {Garg}, Suyog and {Geda}, Robel and {Glattly}, Lauren and {Gondhalekar}, Yash and {Gordon}, Karl D. and {Grant}, David and {Greenfield}, Perry and {Groener}, Austen M. and {Guest}, Steve and {Gurovich}, Sebastian and {Handberg}, Rasmus and {Hart}, Akeem and {Hatfield-Dodds}, Zac and {Homeier}, Derek and {Hosseinzadeh}, Griffin and {Jenness}, Tim and {Jones}, Craig K. and {Joseph}, Prajwel and {Kalmbach}, J. Bryce and {Karamehmetoglu}, Emir and {Ka{l}uszy{'n}ski}, Miko{l}aj and {Kelley}, Michael S.~P. and {Kern}, Nicholas and {Kerzendorf}, Wolfgang E. and {Koch}, Eric W. and {Kulumani}, Shankar and {Lee}, Antony and {Ly}, Chun and {Ma}, Zhiyuan and {MacBride}, Conor and {Maljaars}, Jakob M. and {Muna}, Demitri and {Murphy}, N.~A. and {Norman}, Henrik and {O'Steen}, Richard and {Oman}, Kyle A. and {Pacifici}, Camilla and {Pascual}, Sergio and {Pascual-Granado}, J. and {Patil}, Rohit R. and {Perren}, Gabriel I. and {Pickering}, Timothy E. and {Rastogi}, Tanuj and {Roulston}, Benjamin R. and {Ryan}, Daniel F. and {Rykoff}, Eli S. and {Sabater}, Jose and {Sakurikar}, Parikshit and {Salgado}, Jes{'u}s and {Sanghi}, Aniket and {Saunders}, Nicholas and {Savchenko}, Volodymyr and {Schwardt}, Ludwig and {Seifert-Eckert}, Michael and {Shih}, Albert Y. and {Jain}, Anany Shrey and {Shukla}, Gyanendra and {Sick}, Jonathan and {Simpson}, Chris and {Singanamalla}, Sudheesh and {Singer}, Leo P. and {Singhal}, Jaladh and {Sinha}, Manodeep and {Sip{H{o}}cz}, Brigitta M. and {Spitler}, Lee R. and {Stansby}, David and {Streicher}, Ole and {{{S}}umak}, Jani and {Swinbank}, John D. and {Taranu}, Dan S. and {Tewary}, Nikita and {Tremblay}, Grant R. and {Val-Borro}, Miguel de and {Van Kooten}, Samuel J. and {Vasovi{'c}}, Zlatan and {Verma}, Shresth and {de Miranda Cardoso}, Jos{'e} Vin{'i}cius and {Williams}, Peter K.~G. and {Wilson}, Tom J. and {Winkel}, Benjamin and {Wood-Vasey}, W.~M. and {Xue}, Rui and {Yoachim}, Peter and {Zhang}, Chen and {Zonca}, Andrea and {Astropy Project Contributors}},
        title = "{The Astropy Project: Sustaining and Growing a Community-oriented Open-source Project and the Latest Major Release (v5.0) of the Core Package}",
      journal = {\apj},
         year = 2022,
        month = aug,
       volume = {935},
       number = {2},
          eid = {167},
        pages = {167},
          doi = {10.3847/1538-4357/ac7c74},
archivePrefix = {arXiv},
       eprint = {2206.14220},
 primaryClass = {astro-ph.IM},
       adsurl = {https://ui.adsabs.harvard.edu/abs/2022ApJ...935..167A}
}

@misc{astropy_17756022,
  author       = {Astropy Collaboration},
  title        = {Astropy},
  month        = nov,
  year         = 2025,
  publisher    = {Zenodo},
  version      = {v7.2.0},
  doi          = {10.5281/zenodo.17756022},
  url          = {https://doi.org/10.5281/zenodo.17756022},
}

@ARTICLE{Baibhav:2019,
       author = {{Baibhav}, Vishal and {Berti}, Emanuele and {Gerosa}, Davide and {Mapelli}, Michela and {Giacobbo}, Nicola and {Bouffanais}, Yann and {Di Carlo}, Ugo N.},
        title = "{Gravitational-wave detection rates for compact binaries formed in isolation: {LIGO}/{Virgo} O3 and beyond}",
      journal = {\prd},
         year = 2019,
        month = sep,
       volume = {100},
       number = {6},
          eid = {064060},
        pages = {064060},
          doi = {10.1103/PhysRevD.100.064060},
archivePrefix = {arXiv},
       eprint = {1906.04197},
 primaryClass = {gr-qc},
       adsurl = {https://ui.adsabs.harvard.edu/abs/2019PhRvD.100f4060B}
}

@ARTICLE{Bavera:2020,
       author = {{Bavera}, Simone S. and {Fragos}, Tassos and {Zevin}, Michael and {Berry}, Christopher P.~L. and {Marchant}, Pablo and {Andrews}, Jeff J. and {Coughlin}, Scott and {Dotter}, Aaron and {Kovlakas}, Konstantinos and {Misra}, Devina and {Serra-Perez}, Juan G. and {Qin}, Ying and {Rocha}, Kyle A. and {Rom{\'a}n-Garza}, Jaime and {Tran}, Nam H. and {Zapartas}, Emmanouil},
        title = "{The impact of mass-transfer physics on the observable properties of field binary black hole populations}",
      journal = {\aap},
         year = 2021,
        month = mar,
       volume = {647},
          eid = {A153},
        pages = {A153},
          doi = {10.1051/0004-6361/202039804},
archivePrefix = {arXiv},
       eprint = {2010.16333},
 primaryClass = {astro-ph.HE},
       adsurl = {https://ui.adsabs.harvard.edu/abs/2021A&A...647A.153B}
}

@ARTICLE{Belczynski:2017,
      author = {{Belczynski}, K. and {Askar}, A. and {Arca-Sedda}, M. and {Chruslinska}, M. and {Donnari}, M. and {Giersz}, M. and {Benacquista}, M. and {Spurzem}, R. and {Jin}, D. and {Wiktorowicz}, G. and {Belloni}, D.},
        title = "{The origin of the first neutron star - neutron star merger}",
      journal = {\aap},
         year = 2018,
        month = jul,
       volume = {615},
          eid = {A91},
        pages = {A91},
          doi = {10.1051/0004-6361/201732428},
archivePrefix = {arXiv},
       eprint = {1712.00632},
 primaryClass = {astro-ph.HE},
       adsurl = {https://ui.adsabs.harvard.edu/abs/2018A&A...615A..91B}
}

@ARTICLE{Belczynski:2020,
       author = {{Belczynski}, K. and {Klencki}, J. and {Fields}, C.~E. and {Olejak}, A. and {Berti}, E. and {Meynet}, G. and {Fryer}, C.~L. and {Holz}, D.~E. and {O'Shaughnessy}, R. and others},
        title = "{Evolutionary roads leading to low effective spins, high black hole masses, and O1/O2 rates for {LIGO}/{Virgo} binary black holes}",
      journal = {\aap},
         year = 2020,
        month = apr,
       volume = {636},
          eid = {A104},
        pages = {A104},
          doi = {10.1051/0004-6361/201936528},
archivePrefix = {arXiv},
       eprint = {1706.07053},
 primaryClass = {astro-ph.HE},
       adsurl = {https://ui.adsabs.harvard.edu/abs/2020A&A...636A.104B}
}

@ARTICLE{Boco:2019,
      author = {{Boco}, L. and {Lapi}, A. and {Goswami}, S. and {Perrotta}, F. and {Baccigalupi}, C. and {Danese}, L.},
        title = "{Merging Rates of Compact Binaries in Galaxies: Perspectives for Gravitational Wave Detections}",
      journal = {\apj},
         year = 2019,
        month = aug,
       volume = {881},
       number = {2},
          eid = {157},
        pages = {157},
          doi = {10.3847/1538-4357/ab328e},
archivePrefix = {arXiv},
       eprint = {1907.06841},
 primaryClass = {astro-ph.GA},
       adsurl = {https://ui.adsabs.harvard.edu/abs/2019ApJ...881..157B}
}

@ARTICLE{Boco:2026,
       author = {{Boco}, Lumen and {Bosi}, Michele and {Sgalletta}, Cecilia and {Romagnolo}, Amedeo and {Mapelli}, Michela},
        title = "{Can current models predict the local black hole merger rate?}",
      journal = {arXiv e-prints},
         year = 2026,
        month = jun,
          eid = {arXiv:2606.02725},
        pages = {arXiv:2606.02725},
          doi = {10.48550/arXiv.2606.02725},
archivePrefix = {arXiv},
       eprint = {2606.02725},
 primaryClass = {astro-ph.HE},
       adsurl = {https://ui.adsabs.harvard.edu/abs/2026arXiv260602725B}
}

@ARTICLE{Boesky:2024gw,
       author = {{Boesky}, Adam and {Broekgaarden}, Floor S. and {Berger}, Edo},
        title = "{The Binary Black Hole Merger Rate Deviates From the Cosmic Star Formation Rate: A Tug of War Between Metallicity and Delay Times}",
      journal = {arXiv e-prints},
         year = 2024,
        month = may,
          eid = {arXiv:2405.01623},
        pages = {arXiv:2405.01623},
          doi = {10.48550/arXiv.2405.01623},
archivePrefix = {arXiv},
       eprint = {2405.01623},
 primaryClass = {astro-ph.HE},
       adsurl = {https://ui.adsabs.harvard.edu/abs/2024arXiv240501623B}
}

@ARTICLE{Boesky:2024popsynth,
       author = {{Boesky}, Adam and {Broekgaarden}, Floor S. and {Berger}, Edo},
        title = "{Investigating the Cosmological Rate of Compact Object Mergers from Isolated Massive Binary Stars}",
      journal = {arXiv e-prints},
         year = 2024,
        month = may,
          eid = {arXiv:2405.01630},
        pages = {arXiv:2405.01630},
          doi = {10.48550/arXiv.2405.01630},
archivePrefix = {arXiv},
       eprint = {2405.01630},
 primaryClass = {astro-ph.HE},
       adsurl = {https://ui.adsabs.harvard.edu/abs/2024arXiv240501630B}
}

@ARTICLE{Bouffanais:2021,
       author = {{Bouffanais}, Yann and {Mapelli}, Michela and {Santoliquido}, Filippo and {Giacobbo}, Nicola and {Di Carlo}, Ugo N. and {Rastello}, Sara and {Artale}, M. Celeste and {Iorio}, Giuliano},
        title = "{New insights on binary black hole formation channels after GWTC-2: young star clusters versus isolated binaries}",
      journal = {\mnras},
         year = 2021,
        month = nov,
       volume = {507},
       number = {4},
        pages = {5224-5235},
          doi = {10.1093/mnras/stab2438},
archivePrefix = {arXiv},
       eprint = {2102.12495},
 primaryClass = {astro-ph.HE},
       adsurl = {https://ui.adsabs.harvard.edu/abs/2021MNRAS.507.5224B}
}

@ARTICLE{Breivik:2025review,
       author = {{Breivik}, Katelyn},
        title = "{Population Synthesis of Gravitational Wave Sources}",
      journal = {arXiv e-prints},
         year = 2025,
        month = feb,
          eid = {arXiv:2502.03523},
        pages = {arXiv:2502.03523},
          doi = {10.48550/arXiv.2502.03523},
archivePrefix = {arXiv},
       eprint = {2502.03523},
 primaryClass = {astro-ph.HE},
       adsurl = {https://ui.adsabs.harvard.edu/abs/2025arXiv250203523B}
}

@ARTICLE{Briel:2022,
       author = {{Briel}, M.~M. and {Stevance}, H.~F. and {Eldridge}, J.~J.},
        title = "{Understanding the high-mass binary black hole population from stable mass transfer and super-Eddington accretion in BPASS}",
      journal = {\mnras},
         year = 2023,
        month = apr,
       volume = {520},
       number = {4},
        pages = {5724-5745},
          doi = {10.1093/mnras/stad399},
archivePrefix = {arXiv},
       eprint = {2206.13842},
 primaryClass = {astro-ph.HE},
       adsurl = {https://ui.adsabs.harvard.edu/abs/2023MNRAS.520.5724B}
}

@ARTICLE{Broekgaarden:2021,
       author = {{Broekgaarden}, Floor S. and {Berger}, Edo and {Neijssel}, Coenraad J. and {Vigna-G{\'o}mez}, Alejandro and {Chattopadhyay}, Debatri and {Stevenson}, Simon and {Chruslinska}, Martyna and {Justham}, Stephen and {de Mink}, Selma E. and {Mandel}, Ilya},
        title = "{Impact of massive binary star and cosmic evolution on gravitational wave observations I: black hole-neutron star mergers}",
      journal = {\mnras},
         year = 2021,
        month = dec,
       volume = {508},
       number = {4},
        pages = {5028-5063},
          doi = {10.1093/mnras/stab2716},
archivePrefix = {arXiv},
       eprint = {2103.02608},
 primaryClass = {astro-ph.HE},
       adsurl = {https://ui.adsabs.harvard.edu/abs/2021MNRAS.508.5028B}
}

@ARTICLE{Broekgaarden:2022,
       author = {{Broekgaarden}, Floor S. and {Stevenson}, Simon and {Thrane}, Eric},
        title = "{Signatures of mass ratio reversal in gravitational waves from merging binary black holes}",
      journal = {arXiv e-prints},
         year = 2022,
        month = may,
          eid = {arXiv:2205.01693},
        pages = {arXiv:2205.01693},
archivePrefix = {arXiv},
       eprint = {2205.01693},
 primaryClass = {astro-ph.HE},
       adsurl = {https://ui.adsabs.harvard.edu/abs/2022arXiv220501693B}
}

@dataset{BroekgaardenZenodo:2026-Lower-Your-Rates,
  author       = {Broekgaarden, Floor},
  title        = {Code and data for: Lower Your Rates: On Claims of
                   a Binary Black Hole Merger-Rate Crisis
                  },
  month        = jun,
  year         = 2026,
  publisher    = {Zenodo},
  version      = {version 1},
  doi          = {10.5281/zenodo.20936612},
  url          = {https://doi.org/10.5281/zenodo.20936612},
}

@ARTICLE{Chattaraj:2026,
       author = {{Chattaraj}, Abhishek and {Andrews}, Jeff J. and {Bavera}, Simone S. and {Briel}, Max and {Chattopadhyay}, Debatri and {Fragos}, Tassos and {Gossage}, Seth and {Kalogera}, Vicky and {Kovlakas}, Konstantinos and {Kruckow}, Matthias U. and {Liotine}, Camille and {Rocha}, Kyle A. and {Srivastava}, Philipp M. and {Sun}, Meng and {Teng}, Elizabeth and {Xing}, Zepei and {Zapartas}, Emmanouil},
        title = "{Forming Double Neutron Stars Using Detailed Binary Evolution Models with POSYDON: Comparison to the Galactic Systems}",
      journal = {\apj},
         year = 2026,
        month = jan,
       volume = {997},
       number = {1},
          eid = {52},
        pages = {52},
          doi = {10.3847/1538-4357/ae1f93},
archivePrefix = {arXiv},
       eprint = {2508.00186},
 primaryClass = {astro-ph.SR},
       adsurl = {https://ui.adsabs.harvard.edu/abs/2026ApJ...997...52C}
}

@ARTICLE{Chen:2025,
       author = {{Chen}, Zhiwei and {Zhang}, Jihui and {Lu}, Youjun and {Liu}, Jifeng and {Zeng}, Changwen},
        title = "{Constraining common envelope evolution in binary neutron star formation with combined galactic and gravitational-wave observations}",
      journal = {\mnras},
         year = 2025,
        month = nov,
       volume = {544},
       number = {1},
        pages = {L89-L95},
          doi = {10.1093/mnrasl/slaf105},
archivePrefix = {arXiv},
       eprint = {2508.14397},
 primaryClass = {astro-ph.HE},
       adsurl = {https://ui.adsabs.harvard.edu/abs/2025MNRAS.544L..89C}
}

@ARTICLE{Chruslinska:2018,
       author = {{Chru{\'s}li{\'n}ska}, Martyna and {Belczynski}, Krzysztof and {Klencki}, Jakub and
         {Benacquista}, Matthew},
        title = "{Double neutron stars: merger rates revisited}",
      journal = {\mnras},
         year = 2018,
        month = mar,
       volume = {474},
       number = {3},
        pages = {2937-2958},
          doi = {10.1093/mnras/stx2923},
archivePrefix = {arXiv},
       eprint = {1708.07885},
 primaryClass = {astro-ph.HE},
       adsurl = {https://ui.adsabs.harvard.edu/abs/2018MNRAS.474.2937C}
}

@ARTICLE{Chruslinska:2019,
   author = {{Chruslinska}, M. and {Nelemans}, G. and {Belczynski}, K.},
    title = "{The influence of the distribution of cosmic star formation at different metallicities on the properties of merging double compact objects}",
  journal = {\mnras},
archivePrefix = "arXiv",
   eprint = {1811.03565},
 primaryClass = "astro-ph.HE",
     year = 2019,
    month = feb,
   volume = 482,
    pages = {5012-5017},
      doi = {10.1093/mnras/sty3087},
   adsurl = {https://ui.adsabs.harvard.edu/abs/2019MNRAS.482.5012C}
}

@ARTICLE{Chu:2021,
       author = {{Chu}, Qingbo and {Yu}, Shenghua and {Lu}, Youjun},
        title = "{Formation and Evolution of Binary Neutron Stars: Mergers and Their Host Galaxies}",
      journal = {\mnras},
         year = 2021,
        month = oct,
          doi = {10.1093/mnras/stab2882},
archivePrefix = {arXiv},
       eprint = {2110.04687},
 primaryClass = {astro-ph.GA},
       adsurl = {https://ui.adsabs.harvard.edu/abs/2021MNRAS.tmp.2657C}
}

@ARTICLE{deMinkBelczynski:2015,
       author = {{\noopsort{De Mink}}{de Mink}, S.~E. and {Belczynski}, K.},
        title = "{Merger Rates of Double Neutron Stars and Stellar Origin Black Holes: The Impact of Initial Conditions on Binary Evolution Predictions}",
      journal = {\apj},
         year = "2015",
        month = "Nov",
       volume = {814},
       number = {1},
          eid = {58},
        pages = {58},
          doi = {10.1088/0004-637X/814/1/58},
archivePrefix = {arXiv},
       eprint = {1506.03573},
 primaryClass = {astro-ph.HE},
       adsurl = {https://ui.adsabs.harvard.edu/abs/2015ApJ...814...58D}
}

@ARTICLE{Deng:2024,
       author = {{Deng}, Zhu-Ling and {Li}, Xiang-Dong and {Shao}, Yong and {Xu}, Kun},
        title = "{On the Formation of Double Neutron Stars in the Milky Way: Influence of Key Parameters}",
      journal = {\apj},
         year = 2024,
        month = mar,
       volume = {963},
       number = {2},
          eid = {80},
        pages = {80},
          doi = {10.3847/1538-4357/ad2357},
archivePrefix = {arXiv},
       eprint = {2402.04658},
 primaryClass = {astro-ph.HE},
       adsurl = {https://ui.adsabs.harvard.edu/abs/2024ApJ...963...80D}
}

@ARTICLE{deSa:2024initial,
       author = {{De S{\'a}}, L.~M. and {Rocha}, L.~S. and {Bernardo}, A. and {Bachega}, R.~R.~A. and {Horvath}, J.~E.},
        title = "{Compact object populations over cosmic time II. Compact object merger rates and masses over redshift from varying initial conditions}",
      journal = {arXiv e-prints},
         year = 2024,
        month = oct,
          eid = {arXiv:2410.01451},
        pages = {arXiv:2410.01451},
          doi = {10.48550/arXiv.2410.01451},
archivePrefix = {arXiv},
       eprint = {2410.01451},
 primaryClass = {astro-ph.HE},
       adsurl = {https://ui.adsabs.harvard.edu/abs/2024arXiv241001451D}
}

@ARTICLE{DeSantis:2026,
       author = {{De Santis}, Alessio Ludovico and {Ronchini}, Samuele and {Santoliquido}, Filippo and {Branchesi}, Marica},
        title = "{Constraining Binary Neutron Star Populations using Short Gamma-Ray Burst Observations}",
      journal = {arXiv e-prints},
         year = 2026,
        month = feb,
          eid = {arXiv:2602.13391},
        pages = {arXiv:2602.13391},
          doi = {10.48550/arXiv.2602.13391},
archivePrefix = {arXiv},
       eprint = {2602.13391},
 primaryClass = {astro-ph.HE},
       adsurl = {https://ui.adsabs.harvard.edu/abs/2026arXiv260213391D}
}

@ARTICLE{Dominik:2014,
   author = {{Dominik}, M. and {Berti}, E. and {O'Shaughnessy}, R. and {Mandel}, I. and 
	{Belczynski}, K. and {Fryer}, C. and {Holz}, D.~E. and {Bulik}, T. and 
	{Pannarale}, F.},
    title = "{Double Compact Objects III: Gravitational-wave Detection Rates}",
  journal = {\apj},
archivePrefix = "arXiv",
   eprint = {1405.7016},
 primaryClass = "astro-ph.HE",
     year = 2015,
    month = jun,
   volume = 806,
      eid = {263},
    pages = {263},
      doi = {10.1088/0004-637X/806/2/263},
   adsurl = {http://adsabs.harvard.edu/abs/2015ApJ...806..263D}
}

@ARTICLE{DorozsmaiToonen:2022,
       author = {{Dorozsmai}, Andris and {Toonen}, Silvia},
        title = "{Importance of stable mass transfer and stellar winds for the formation of gravitational wave sources}",
      journal = {\mnras},
         year = 2024,
        month = jun,
       volume = {530},
       number = {4},
        pages = {3706-3739},
          doi = {10.1093/mnras/stae152},
archivePrefix = {arXiv},
       eprint = {2207.08837},
 primaryClass = {astro-ph.SR},
       adsurl = {https://ui.adsabs.harvard.edu/abs/2024MNRAS.530.3706D}
}

@ARTICLE{Eldridge:2019,
       author = {{Eldridge}, J.~J. and {Stanway}, E.~R. and {Tang}, Petra N.},
        title = "{A consistent estimate for gravitational wave and electromagnetic transient rates}",
      journal = {\mnras},
         year = 2019,
        month = jan,
       volume = {482},
       number = {1},
        pages = {870-880},
          doi = {10.1093/mnras/sty2714},
archivePrefix = {arXiv},
       eprint = {1807.07659},
 primaryClass = {astro-ph.HE},
       adsurl = {https://ui.adsabs.harvard.edu/abs/2019MNRAS.482..870E}
}

@ARTICLE{Ghodla:2021,
       author = {{Ghodla}, Sohan and {van Zeist}, Wouter G.~J. and {Eldridge}, J.~J. and {Stevance}, H{\'e}lo{\"\i}se F. and {Stanway}, Elizabeth R.},
        title = "{Forward Modelling the {LIGO}/VIRGO O3a GW transient mass distributions with BPASS}",
      journal = {arXiv e-prints},
         year = 2021,
        month = may,
archivePrefix = {arXiv},
       eprint = {2105.05783},
 primaryClass = {astro-ph.HE},
       adsurl = {https://ui.adsabs.harvard.edu/abs/2021arXiv210505783G}
}

@ARTICLE{GiacobboMapelli:2018,
       author = {{Giacobbo}, Nicola and {Mapelli}, Michela},
        title = "{The progenitors of compact-object binaries: impact of metallicity, common envelope and natal kicks}",
      journal = {\mnras},
         year = "2018",
        month = "Oct",
       volume = {480},
       number = {2},
        pages = {2011-2030},
          doi = {10.1093/mnras/sty1999},
archivePrefix = {arXiv},
       eprint = {1806.00001},
 primaryClass = {astro-ph.HE},
       adsurl = {https://ui.adsabs.harvard.edu/abs/2018MNRAS.480.2011G}
}

@ARTICLE{GiacobboMapelli:2020,
       author = {{Giacobbo}, Nicola and {Mapelli}, Michela},
        title = "{Revising Natal Kick Prescriptions in Population Synthesis Simulations}",
      journal = {\apj},
         year = 2020,
        month = mar,
       volume = {891},
       number = {2},
          eid = {141},
        pages = {141},
          doi = {10.3847/1538-4357/ab7335},
archivePrefix = {arXiv},
       eprint = {1909.06385},
 primaryClass = {astro-ph.HE},
       adsurl = {https://ui.adsabs.harvard.edu/abs/2020ApJ...891..141G}
}

@ARTICLE{GWTC-4-pop,
       author = {{The LIGO Scientific Collaboration} and {the Virgo Collaboration} and {the KAGRA Collaboration} and {Abac}, A.~G. and {Abouelfettouh}, I. and {Acernese}, F. and {Ackley}, K. and {Adamcewicz}, C. and {Adhicary}, S. and {Adhikari}, D. and {Adhikari}, N. and {Adhikari}, R.~X. and {Adkins}, V.~K. and {Afroz}, S. and {Agarwal}, D. and {Agathos}, M. and {Aghaei Abchouyeh}, M. and {Aguiar}, O.~D. and {Ahmadzadeh}, S. and {Aiello}, L. and {Ain}, A. and {Ajith}, P. and {Akutsu}, T. and {Albanesi}, S. and {Alfaidi}, R.~A. and {Al-Jodah}, A. and {All{\'e}n{\'e}}, C. and {Allocca}, A. and {Al-Shammari}, S. and {Altin}, P.~A. and {Alvarez-Lopez}, S. and {Amarasinghe}, O. and {Amato}, A. and {Amra}, C. and {Ananyeva}, A. and {Anderson}, S.~B. and {Anderson}, W.~G. and {Andia}, M. and {Ando}, M. and {Andrade}, T. and {Andr{\'e}s-Carcasona}, M. and {Andri{\'c}}, T. and {Anglin}, J. and {Ansoldi}, S. and {Antelis}, J.~M. and {Antier}, S. and {Aoumi}, M. and {Appavuravther}, E.~Z. and {Appert}, S. and {Apple}, S.~K. and {Arai}, K. and {Araya}, A. and {Araya}, M.~C. and {Arca Sedda}, M. and {Areeda}, J.~S. and {Argianas}, L. and {Aritomi}, N. and {Armato}, F. and {Armstrong}, S. and {Arnaud}, N. and {Arogeti}, M. and {Aronson}, S.~M. and {Arun}, K.~G. and {Ashton}, G. and {Aso}, Y. and {Assiduo}, M. and {Assis de Souza Melo}, S. and {Aston}, S.~M. and {Astone}, P. and {Attadio}, F. and {Aubin}, F. and {AultONeal}, K. and {Avallone}, G. and {Babak}, S. and {Badaracco}, F. and {Badger}, C. and {Bae}, S. and {Bagnasco}, S. and {Bagui}, E. and {Baiotti}, L. and {Bajpai}, R. and {Baka}, T. and {Baker}, T. and {Ball}, M. and {Ballardin}, G. and {Ballmer}, S.~W. and {Banagiri}, S. and {Banerjee}, B. and {Bankar}, D. and {Baptiste}, T.~M. and {Baral}, P. and {Barayoga}, J.~C. and {Barish}, B.~C. and {Barker}, D. and {Barman}, N. and {Barneo}, P. and {Barone}, F. and {Barr}, B. and {Barsotti}, L. and {Barsuglia}, M. and {Barta}, D. and {Bartoletti}, A.~M. and {Barton}, M.~A. and {Bartos}, I. and {Basak}, S. and {Basalaev}, A. and {Bassiri}, R. and {Basti}, A. and {Bates}, D.~E. and {Bawaj}, M. and {Baxi}, P. and {Bayley}, J.~C. and {Baylor}, A.~C. and {Baynard}, II, P.~A. and {Bazzan}, M. and {Bedakihale}, V.~M. and {Beirnaert}, F. and {Bejger}, M. and {Belardinelli}, D. and {Bell}, A.~S. and {Bellie}, D.~S. and {Bellizzi}, L. and {Beltran-Martinez}, D. and {Benoit}, W. and {Bentara}, I. and {Bentley}, J.~D. and {Ben Yaala}, M. and {Bera}, S. and {Bergamin}, F. and {Berger}, B.~K. and {Bernuzzi}, S. and {Beroiz}, M. and {Berry}, C.~P.~L. and {Bersanetti}, D. and {Bertolini}, A. and {Betzwieser}, J. and {Beveridge}, D. and {Bevilacqua}, G. and {Bevins}, N. and {Bhandare}, R. and {Bhatt}, R. and {Bhattacharjee}, D. and {Bhaumik}, S. and {Bhowmick}, S. and {Biancalana}, V. and {Bianchi}, A. and {Bilenko}, I.~A. and {Billingsley}, G. and {Binetti}, A. and {Bini}, S. and {Binu}, C. and {Birnholtz}, O. and {Biscoveanu}, S. and {Bisht}, A. and {Bitossi}, M. and {Bizouard}, M. -A. and {Blaber}, S. and {Blackburn}, J.~K. and {Blagg}, L.~A. and {Blair}, C.~D. and {Blair}, D.~G. and {Bobba}, F. and {Bode}, N. and {Boileau}, G. and {Boldrini}, M. and {Bolingbroke}, G.~N. and {Bolliand}, A. and {Bonavena}, L.~D. and {Bondarescu}, R. and {Bondu}, F. and {Bonilla}, E. and {Bonilla}, M.~S. and {Bonino}, A. and {Bonnand}, R. and {Booker}, P. and {Borchers}, A. and {Borhanian}, S. and {Boschi}, V. and {Bose}, S. and {Bossilkov}, V. and {Boudon}, A. and {Bozzi}, A. and {Bradaschia}, C. and {Brady}, P.~R. and {Branch}, A. and {Branchesi}, M. and {Braun}, I. and {Briant}, T. and {Brillet}, A. and {Brinkmann}, M. and {Brockill}, P. and {Brockmueller}, E. and {Brooks}, A.~F. and {Brown}, B.~C. and {Brown}, D.~D. and {Brozzetti}, M.~L. and {Brunett}, S. and {Bruno}, G. and {Bruntz}, R. and {Bryant}, J.},
        title = "{GWTC-4.0: Population Properties of Merging Compact Binaries}",
      journal = {arXiv e-prints},
         year = 2025,
        month = aug,
          eid = {arXiv:2508.18083},
        pages = {arXiv:2508.18083},
          doi = {10.48550/arXiv.2508.18083},
archivePrefix = {arXiv},
       eprint = {2508.18083},
 primaryClass = {astro-ph.HE},
       adsurl = {https://ui.adsabs.harvard.edu/abs/2025arXiv250818083T}
}

@ARTICLE{GWTC-5:catalog,
       author = {{The LIGO Scientific Collaboration} and {the Virgo Collaboration} and {the KAGRA Collaboration}},
        title = "{GWTC-5.0: Observations from the Second Part of the Fourth LIGO-Virgo-KAGRA Observing Run and Updates to the Gravitational-Wave Transient Catalog}",
      journal = {arXiv e-prints},
         year = 2026,
        month = may,
          eid = {arXiv:2605.27225},
        pages = {arXiv:2605.27225},
archivePrefix = {arXiv},
       eprint = {2605.27225},
 primaryClass = {gr-qc},
       adsurl = {https://ui.adsabs.harvard.edu/abs/2026arXiv260527225T}
}

@ARTICLE{GWTC-5:populations,
       author = {{The LIGO Scientific Collaboration} and {the Virgo Collaboration} and {the KAGRA Collaboration}},
        title = "{GWTC-5.0: Population Properties of Merging Compact Binaries}",
      journal = {arXiv e-prints},
         year = 2026,
        month = may,
          eid = {arXiv:2605.27226},
        pages = {arXiv:2605.27226},
archivePrefix = {arXiv},
       eprint = {2605.27226},
 primaryClass = {astro-ph.HE},
       adsurl = {https://ui.adsabs.harvard.edu/abs/2026arXiv260527226T}
}

@ARTICLE{GWTC2,
       author = {{Abbott}, R. and {Abbott}, T.~D. and {Abraham}, S. and {Acernese}, F. and
         others},
        title = "{GWTC-2: Compact Binary Coalescences Observed by LIGO and Virgo during the First Half of the Third Observing Run}",
      journal = {Physical Review X},
         year = 2021,
        month = apr,
       volume = {11},
       number = {2},
          eid = {021053},
        pages = {021053},
          doi = {10.1103/PhysRevX.11.021053},
archivePrefix = {arXiv},
       eprint = {2010.14527},
 primaryClass = {gr-qc},
       adsurl = {https://ui.adsabs.harvard.edu/abs/2021PhRvX..11b1053A}
}

@ARTICLE{GWTC3,
       author = {{Abbott}, R. and {et al (LIGO Scientific, Virgo, and KAGRA collaborations)}},
        title = "{GWTC-3: Compact Binary Coalescences Observed by LIGO and Virgo during the Second Part of the Third Observing Run}",
      journal = {Physical Review X},
         year = 2023,
        month = oct,
       volume = {13},
       number = {4},
          eid = {041039},
        pages = {041039},
          doi = {10.1103/PhysRevX.13.041039},
archivePrefix = {arXiv},
       eprint = {2111.03606},
 primaryClass = {gr-qc},
       adsurl = {https://ui.adsabs.harvard.edu/abs/2023PhRvX..13d1039A}
}

@book{python,
  author    = {Van Rossum, Guido and Drake, Fred L.},
  title     = {Python 3 Reference Manual},
  year      = {2009},
  isbn      = {1441412697},
  publisher = {CreateSpace},
  address   = {Scotts Valley, CA}
}

\begin{appendix}
\twocolumngrid

\begin{figure*}
    \centering
    \includegraphics[width=1\linewidth]{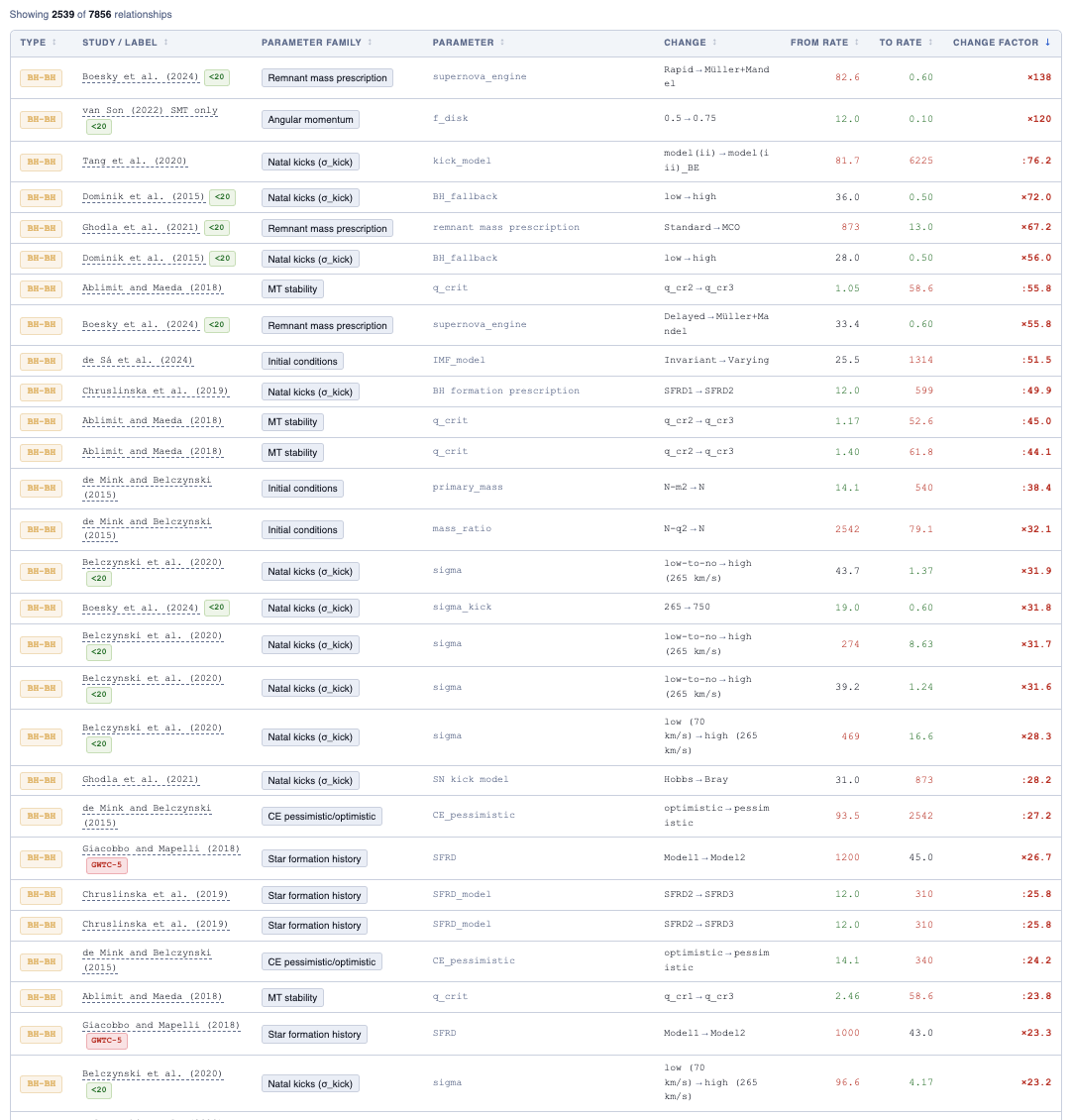}
    \caption{Excerpt from the online interactive table of one-parameter-at-a-time relationships between isolated-binary-evolution submodels, ranked by largest change in BBH rate. 
     Full table available at \citep{BroekgaardenZenodo:2026-Lower-Your-Rates} and \href{https://floorbroekgaarden.github.io/lower-your-rates/}{https://floorbroekgaarden.github.io/lower-your-rates/}.
    Each row connects two submodels from the same study that differ in a single parameter or physical assumption and reports the corresponding change in the simulated  local BBH merger rate density. Relationships are grouped by parameter family and ordered here by decreasing absolute change factor. The `from' and `to' rates are given in $\Gpcyr$, with green values indicating lower rates and red values indicating higher rates. Labels identify submodels that fall below $20\Gpcyr$ or within the GWTC-5 observational interval. The full interactive table contains $2543$ BH--BH relationships and allows users to sort and filter by study, parameter family, parameter, and rate change.
    }
    \label{tab:examples-rate-reductions}
\end{figure*}

\begin{figure*}[t]
    \centering
    \captionsetup[subfigure]{font=small, skip=2pt}

    \begin{subfigure}[t]{0.43\textwidth}
        \centering
        \includegraphics[width=\linewidth]
        {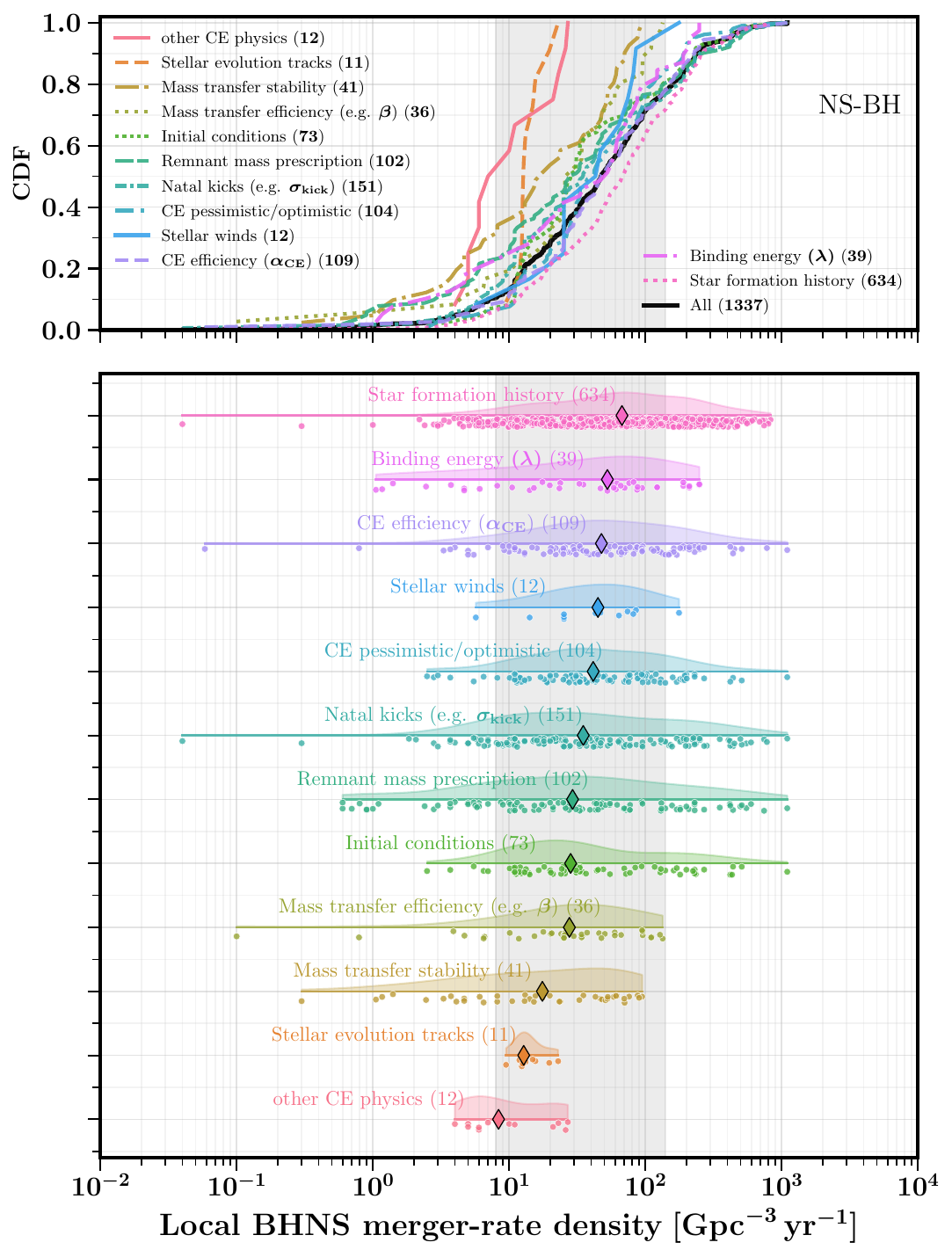}
        \caption{BHNS merger rates grouped by parameter family.}
        \label{fig:parameter-variation-violin_BHNS}
    \end{subfigure}
    \hspace{0.04\textwidth}
    \begin{subfigure}[t]{0.43\textwidth}
        \centering
        \includegraphics[width=\linewidth]
        {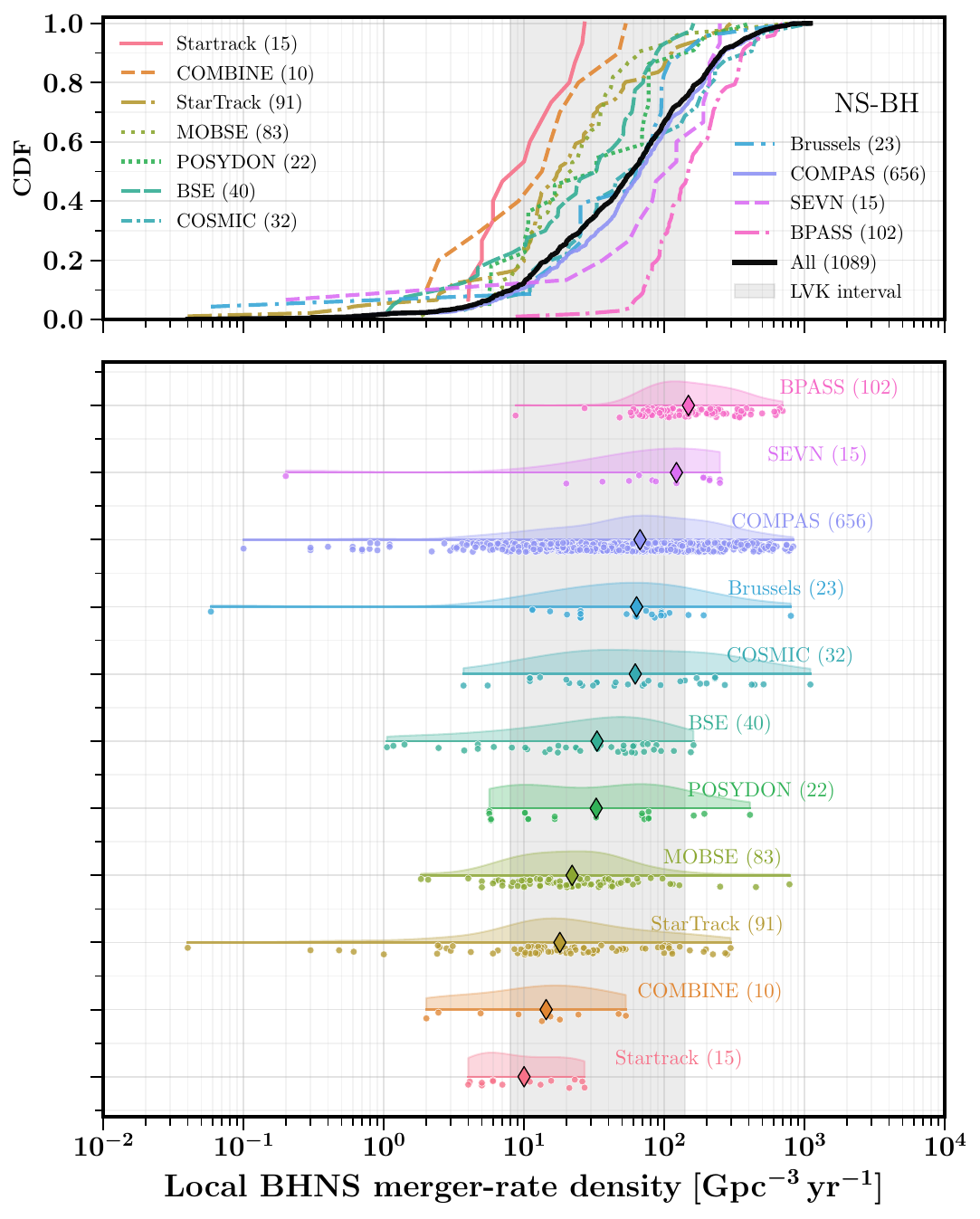}
        \caption{BHNS merger rates grouped by population-synthesis code.}
        \label{fig:population-synthesis-code_BHNS}
    \end{subfigure}

    \vspace{0.15cm}

    \begin{subfigure}[t]{0.43\textwidth}
        \centering
        \includegraphics[width=\linewidth]
        {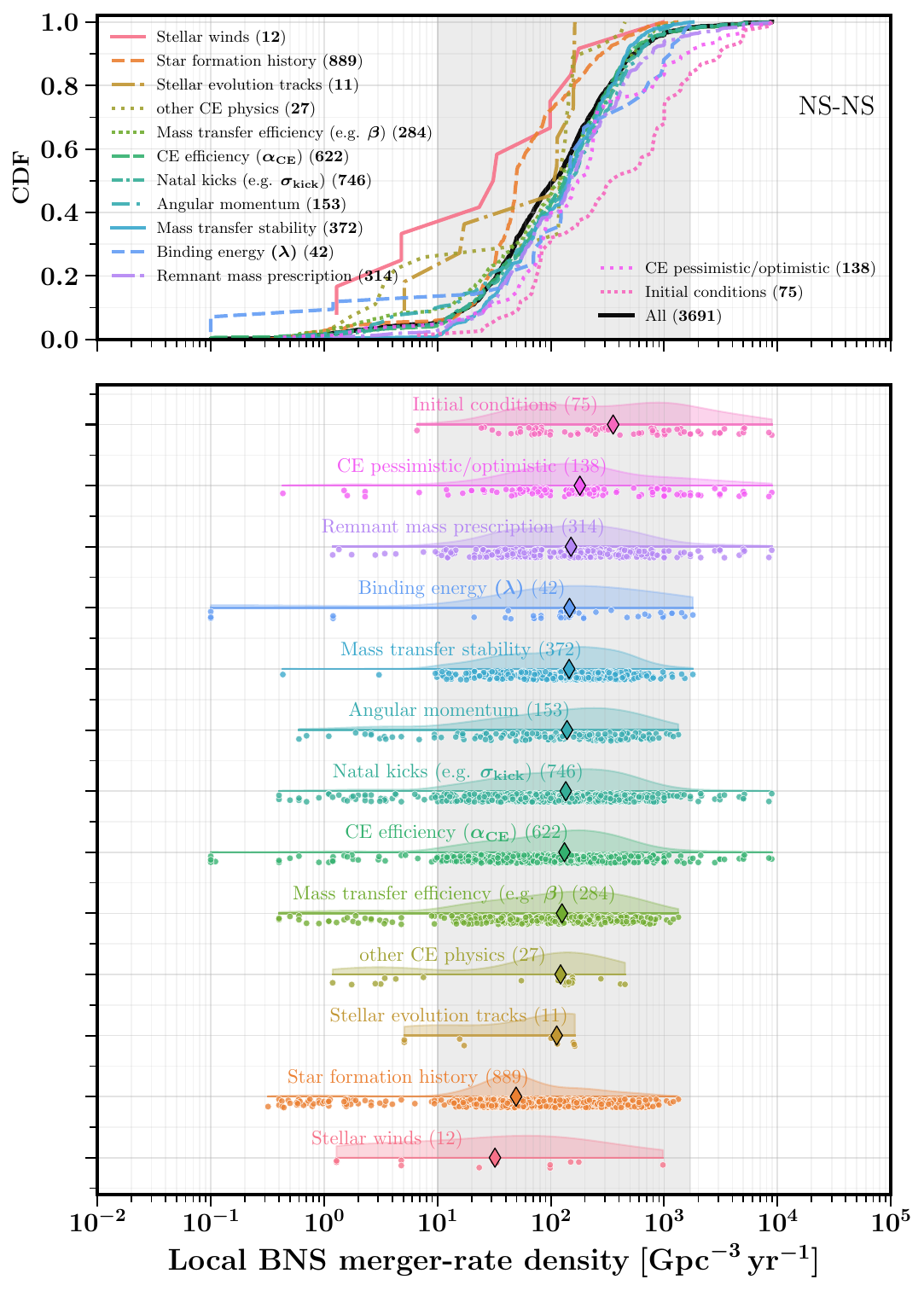}
        \caption{BNS merger rates grouped by parameter family.}
        \label{fig:parameter-variation-violin_BNS}
    \end{subfigure}
    \hspace{0.04\textwidth}
    \begin{subfigure}[t]{0.43\textwidth}
        \centering
        \includegraphics[width=\linewidth]
        {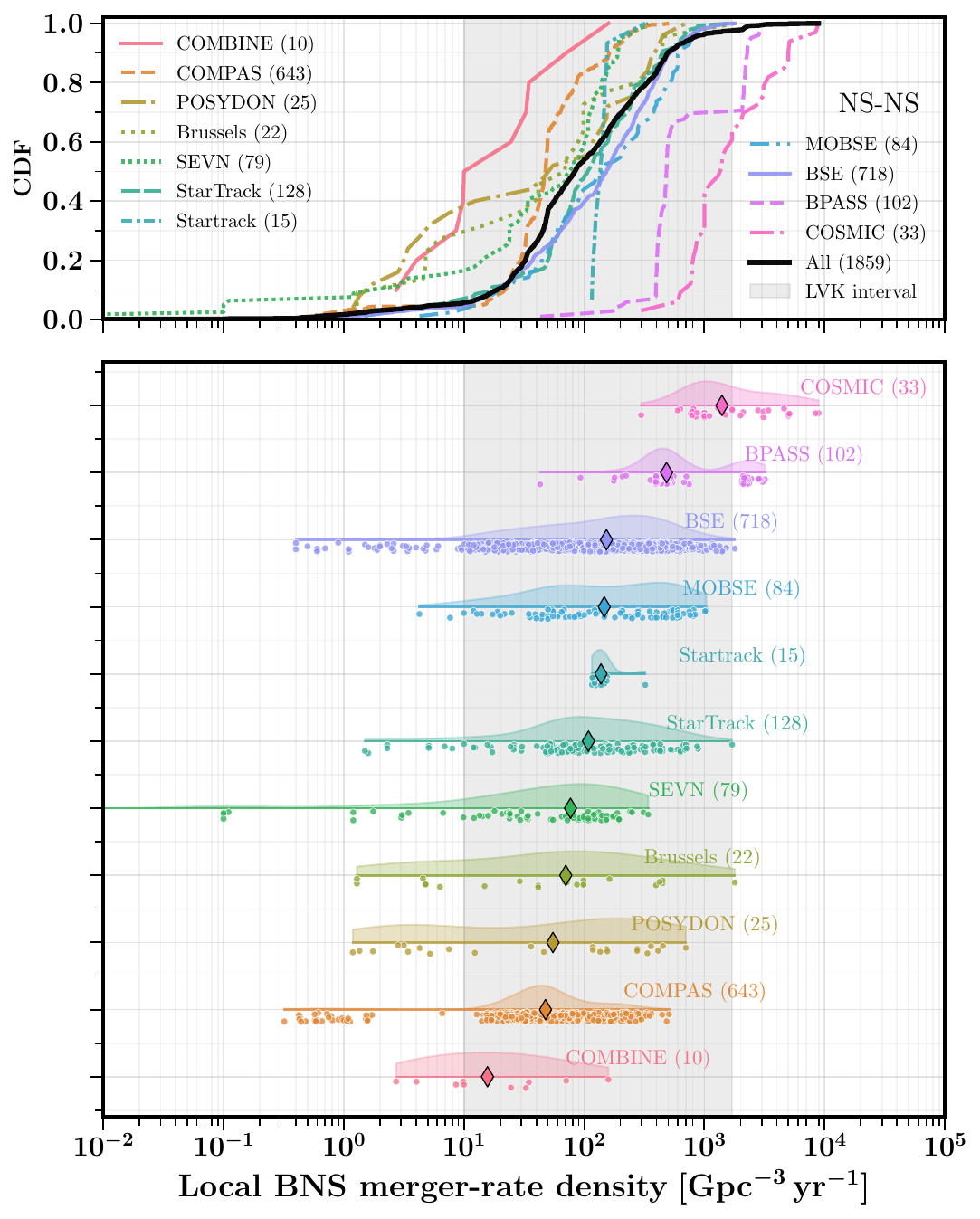}
        \caption{BNS merger rates grouped by population-synthesis code.}
        \label{fig:population-synthesis-code_BNS}
    \end{subfigure}

    \caption{
        Local merger-rate distributions for BHNS and BNS systems.
        Panels (a) and (c) are the same as
        Figure~\ref{fig:parameter-variation-violin}, grouped by parameter
        family, while panels (b) and (d) are the same as
        Figure~\ref{fig:violin-population-synthesis-code}, grouped by
        population-synthesis code.
    }
    \label{fig:BHNS-BNS-rate-comparison}
\end{figure*}

\begin{figure*}
\centering
\includegraphics[width=0.8\linewidth]{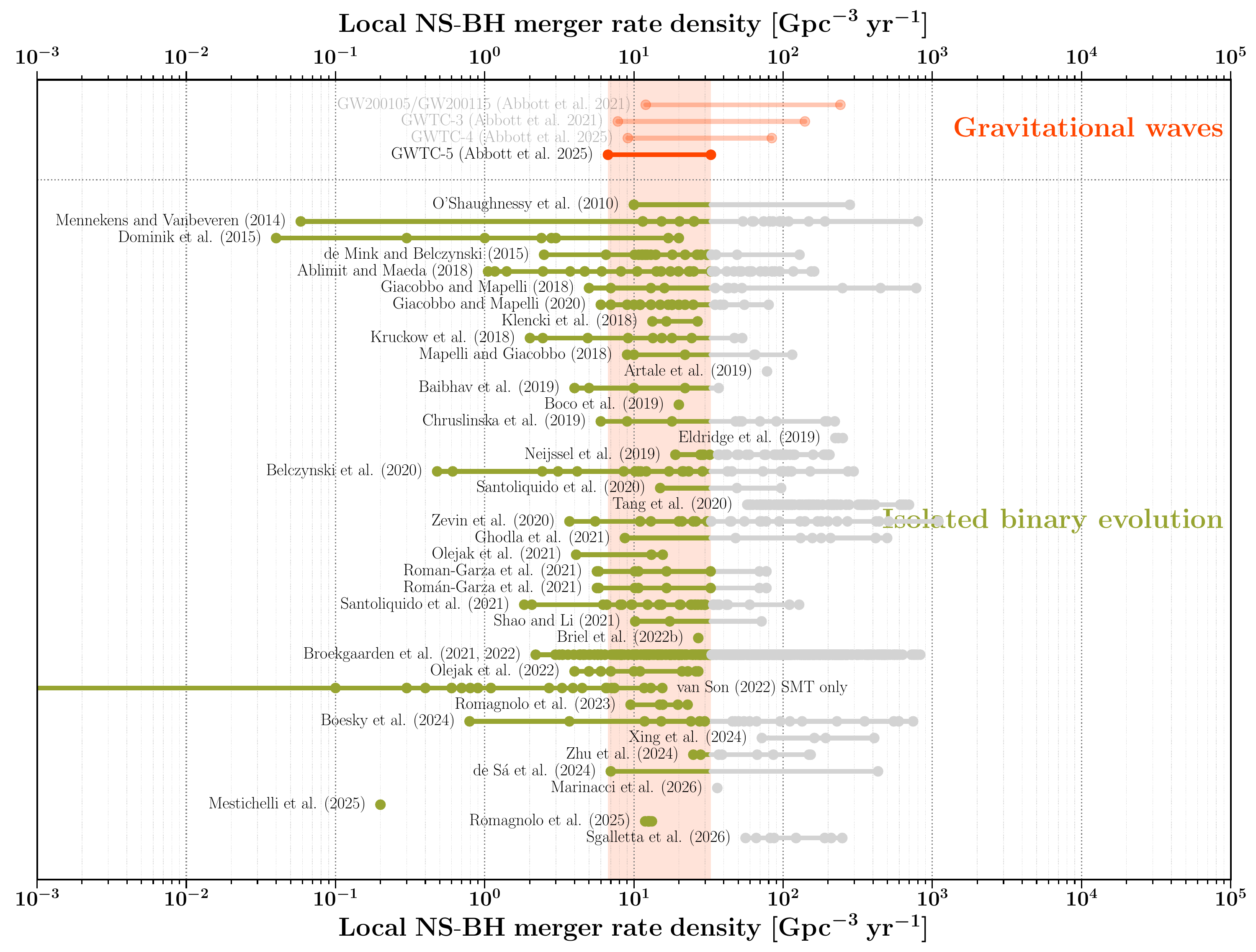}
\caption{
Same as Figure~\ref{fig:BBH-isolated-review-rates} for BHNS.
}
\label{fig:BHNS-isolated-review-rates}
\includegraphics[width=0.8\linewidth]{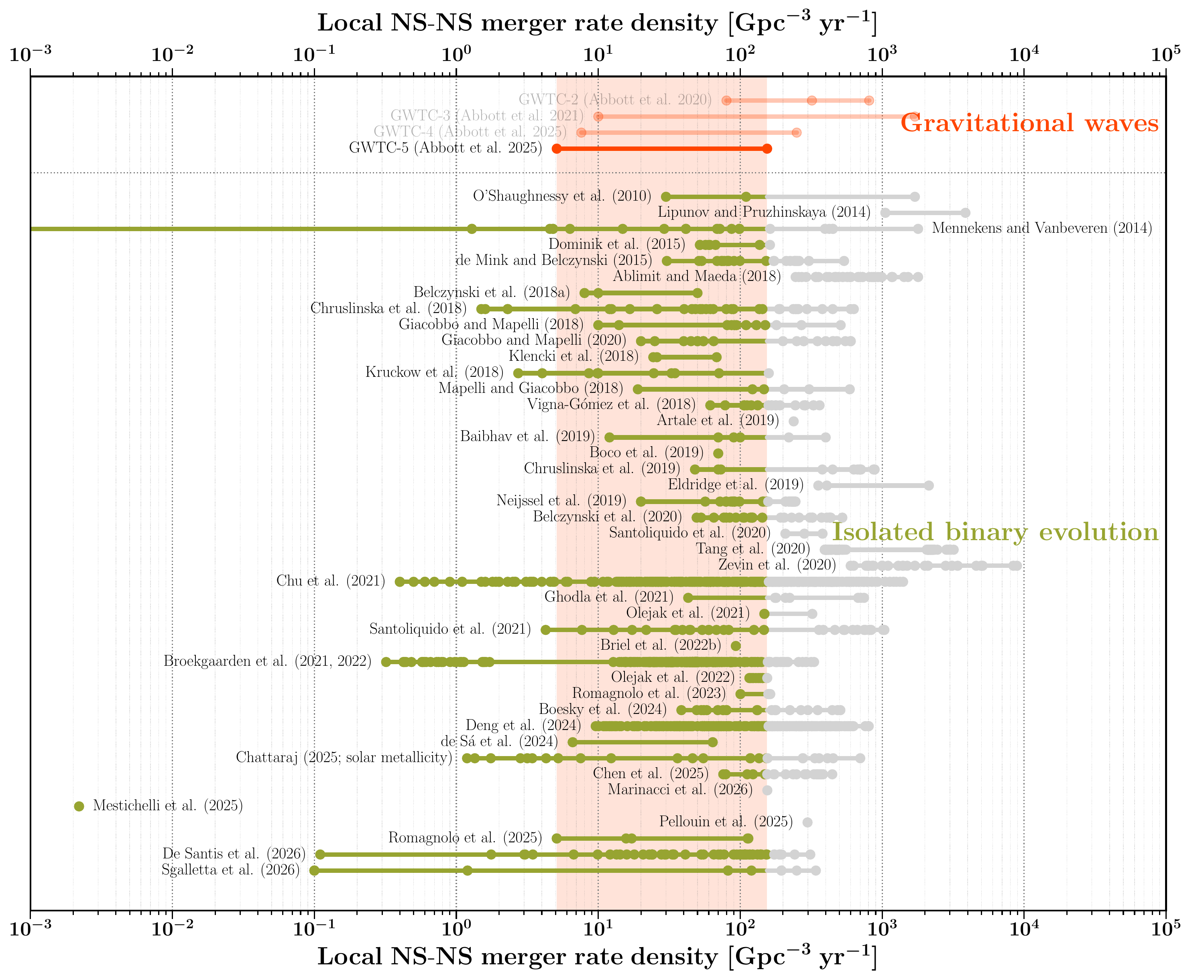}
\caption{
Same as Figure~\ref{fig:BBH-isolated-review-rates} for BNS.
}
\label{fig:BNS-isolated-review-rates}
\end{figure*}

\section{Extra Figures}
\label{ap:ap-BHNS-and-BNS-parameter-variations}

We provide an example exerpt from our online table in Figure~\ref{tab:examples-rate-reductions}.  We also provide  additional figures that show the same analysis for BHNS and BNS mergers in Figures~\ref{fig:parameter-variation-violin_BHNS}, \ref{fig:population-synthesis-code_BHNS},\ref{fig:parameter-variation-violin_BNS}, \ref{fig:population-synthesis-code_BNS}, \ref{fig:BHNS-isolated-review-rates} and \ref{fig:BNS-isolated-review-rates}.

\end{appendix}

\end{document}